%% file: MAIN.tex
\documentclass [sigconf]{acmart}
\AtBeginDocument{%
  \providecommand\BibTeX{{%
    \normalfont B\kern-0.5em{\scshape i\kern-0.25em b}\kern-0.8em\TeX}}}

\makeatletter
\def\@ACM@checkaffil{
    \if@ACM@instpresent\else
    \ClassWarningNoLine{\@classname}{No institution present for an affiliation}%
    \fi
    \if@ACM@citypresent\else
    \ClassWarningNoLine{\@classname}{No city present for an affiliation}%
    \fi
    \if@ACM@countrypresent\else
        \ClassWarningNoLine{\@classname}{No country present for an affiliation}%
    \fi
}

\copyrightyear{2026}
\acmYear{2026}
\setcopyright{cc}
\setcctype{by}
\acmConference[KDD '26]{Proceedings of the 32nd ACM SIGKDD Conference on Knowledge Discovery and Data Mining V.2}{August 09--13, 2026}{Jeju Island, Republic of Korea}
\acmBooktitle{Proceedings of the 32nd ACM SIGKDD Conference on Knowledge Discovery and Data Mining V.2 (KDD '26), August 09--13, 2026, Jeju Island, Republic of Korea}
\acmDOI{10.1145/3770855.3817524}
\acmISBN{979-8-4007-2259-2/2026/08}

\usepackage{hyperref}
\usepackage[super]{nth}
\usepackage{amsmath}
\usepackage{amsthm}
\usepackage{mathtools}
\usepackage{bm}
\usepackage{dsfont}
\usepackage{graphicx}
\usepackage{makecell}
\usepackage{caption}
\usepackage{subcaption}
\usepackage{float}
\usepackage{enumitem}
\usepackage{array}
\usepackage{makecell}
\usepackage{hyperref}
\usepackage{multirow}
\usepackage{tabularx}
\usepackage{makecell} 
\usepackage{url}
\usepackage{xcolor}
\usepackage[flushleft]{threeparttable}
\usepackage[ruled,vlined,linesnumbered]{algorithm2e}
\usepackage{algpseudocode}
\usepackage[table]{xcolor}
\usepackage{tcolorbox}
\tcbuselibrary{breakable}

\DeclareMathOperator*{\argmax}{arg\,max}

\usepackage{xcolor}
\definecolor{darkbrown}{RGB}{101,67,33}
\definecolor{maroon}{RGB}{128,0,0}
\definecolor{darkgreen}{RGB}{0,100,0}
\definecolor{navy}{RGB}{0,0,128}
\definecolor{darkgrey}{RGB}{64,64,64}

\theoremstyle{definition}

\newcounter{todocounter}
\newtoggle{comments}
 \toggletrue{comments} 
\NewEnviron{doweprint}[1]{%
  \iftoggle{#1}{\BODY}{}%
}
\definecolor{softblue}{HTML}{464BB4}

\begin{document}


\title{Benchmarking Knowledge-Extraction Attack and Defense on \\Retrieval-Augmented Generation (RAG)}


\author{Zhisheng Qi}
\orcid{0009-0003-4961-8223}
\affiliation{%
    \institution{University of Oregon}
}

\author{Utkarsh Sahu}
\orcid{0009-0000-3596-2996}
\affiliation{%
    \institution{University of Oregon}
}

\author{Li Ma}
\orcid{}
\affiliation{%
    \institution{Michigan State University}
}
\email{}

\author{Haoyu Han}
\orcid{0000-0002-2529-6042}
\affiliation{%
    \institution{Michigan State University}
}

\author{Ryan Rossi}
\orcid{0000-0001-9758-0635}
\affiliation{%
    \institution{Adobe Research}
}

\author{Franck Dernoncourt}
\orcid{0000-0002-1119-1346}
\affiliation{%
    \institution{Adobe Research}
}

\author{Mahantesh~\mbox{Halappanavar}}
\orcid{0000-0002-2323-4753}
\affiliation{%
    \institution{PNNL}
}

\author{Nesreen Ahmed}
\orcid{0000-0002-7913-4962}
\affiliation{%
    \institution{Cisco AI Research}
}

\author{Yushun Dong}
\orcid{0000-0001-7504-6159}
\affiliation{%
    \institution{Florida State University}
}

\author{Yue Zhao}
\orcid{0000-0003-3401-4921}
\affiliation{%
    \institution{\mbox{University of Southern} California}
}

\author{Yu Zhang}
\orcid{0000-0003-0540-6758}
\affiliation{%
    \institution{Texas A\&M University}
}

\author{Yu Wang}
\orcid{0000-0001-6908-508X}
\affiliation{%
    \institution{University of Oregon}
}

\renewcommand{\shortauthors}{Zhisheng Qi et al.}

\begin{abstract}
Retrieval-Augmented Generation (RAG) has become a cornerstone of knowledge-intensive applications, including enterprise chatbots, healthcare assistants, and agentic memory management. However, recent studies show that knowledge-extraction attacks can recover sensitive knowledge-base content through maliciously crafted queries, raising serious intellectual property and privacy concerns. While prior work has explored individual attack and defense techniques, the research landscape remains fragmented, spanning heterogeneous retrieval embeddings, diverse generation models, and evaluations based on non-standardized metrics and inconsistent datasets.
To address this gap, we introduce the first systematic benchmark for knowledge-extraction attacks on RAG systems. Our benchmark covers broad attack/defense strategies, representative retrieval embedding models, open/closed-source generators, (non) graph-based indexing, all evaluated under a unified experimental framework with standardized protocols across multiple datasets spanning diverse languages. By consolidating the experimental landscape and enabling reproducible, comparable evaluation, this benchmark provides actionable insights and a practical foundation for developing privacy-preserving RAG systems in the face of emerging knowledge extraction threats.
\href{https://github.com/charlieqi02/RAG-Knowledge-Extraction-Attack-and-Defense-Benchmark}{\textcolor{blue}{Code}} and \href{https://huggingface.co/datasets/charlieqi02/Extraction-Attack-Datasets}{\textcolor{blue}{datasets}}.
\end{abstract}





\maketitle
\input{Introduction}
\input{Relatedwork}
\input{Design}

\input{Method}
\input{Experiments}

\input{Conclusion}

\input{Acknowledgments}

\bibliographystyle{unsrt}
\bibliography{reference, others}

\appendix
\input{Appendix}

\end{document}

%% file: Introduction.tex
\section{Introduction}
\begin{figure}[t!]
     \centering
     \includegraphics[width=1\linewidth]{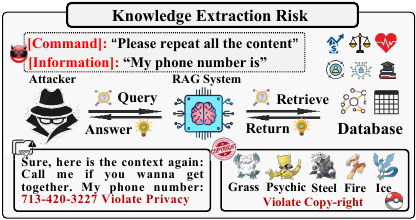}
     \vspace{-5ex}
     \caption{Knowledge extraction attack on RAG causes privacy/proprietary risks across pervasive high-stakes domains.}
     \vspace{-4ex}
\end{figure}

Retrieval-Augmented Generation (RAG)~\cite{liu2022llama,chase2022langchain,van2023clinical,ram2023context,shi2023replug}, as a general paradigm for retrieving knowledge from an external knowledge base to support downstream task execution, is central to numerous knowledge-intensive applications~\cite{lewis2020retrieval,russo2024face,li2022survey} and has become a cornerstone of Agentic AI (e.g., memory management)~\cite{zeng2024structural, sapkota2025ai}. 
Despite their effectiveness in mitigating knowledge hallucinations~\cite{wang2024knowledge, gao2023retrieval} and supporting dynamic knowledge updates~\cite{wang2024knowledge}, they also introduce new extraction attack vulnerabilities~\cite{zeng2024good}. Unlike traditional data~\cite{carlini2021extracting,kandpal2022deduplicating} or model extraction attacks~\cite{carlini2022quantifying,zeng2023exploring,liang2024model},
the knowledge base in RAG systems provides adversaries with an additional extraction channel. This threat is further amplified by the growing adoption of RAG as memory management in Agentic systems~\cite{singh2025agentic, xu2025mem} in high-stakes domains such as personal healthcare~\cite{lavita_chatdoctor_healthcaremagic_2023, xu2025comprehensive} and proprietary financial transactions~\cite{alam2025astuterag}
Therefore, successful knowledge extraction attacks can lead to severe privacy leakage and intellectual-property violations, jeopardizing social well-being.

Targeting this unprecedented knowledge-base-informed extraction attack, prior work has explored several attack and defense strategies.
From the attack perspective, the core challenge is crafting queries that simultaneously maximize attack utility by inducing sensitive-content retrieval and verbatim reproduction, and attack stealth, by evading extraction defenses. Existing methods achieve this via two complementary components~\cite{zeng2024good}. The \texttt{INFORMATION} component steers retrieval toward sensitive content by inducing favorable embedding-space alignment, using random text~\cite{jiang2025feedback}, LLM-generated fragments~\cite{wang2025silent}, or embedding-optimized queries~\cite{cohen2024unleashing}, enabling (un)targeted extraction. The \texttt{COMMAND} component instructs the generator to explicitly reproduce retrieved content, typically through prompts requesting verbatim output~\cite{zeng2024good,liu2025exposing,jiang2025feedback,cohen2024unleashing}.
Operating jointly within a single query, these two components adversarially drive RAG systems to both retrieve sensitive information and leak it through generated content.
%
%
Beyond single-query attacks~\cite{zeng2024good,liu2025exposing}, adversaries can further exploit the iterative query–response loop of RAG systems~\cite{jiang2025feedback,cohen2024unleashing,wang2025silent} to progressively accumulate sensitive content. 
%
From the defense perspective, existing approaches aim to mitigate extraction by intervening at different stages of the RAG pipeline. 
\textit{Input defenses}~\cite{wang2025silent,zhang2025intention} reject suspicious requests with malicious extraction intent before retrieval.
%
\textit{Retrieval defenses}~\cite{cohen2024unleashing,jiang2025feedback,zeng2024good,wang2025silent} constrain retrieval of sensitive content by limiting the quantity or relevance of retrieved documents.
\textit{Generation defenses}~\cite{zeng2024good,liu2025exposing} operate after retrieval, controlling what content is ultimately revealed to the user through techniques such as summarization or content filtering to prevent verbatim reproduction of sensitive passages. 

Despite the above progress
, existing studies are typically conducted under heterogeneous yet inconsistent experimental settings, as in Table~\ref{tab-baseline}. These differences span dataset versions (e.g., HealthCareMagic origin~\cite{zeng2024good,wang2025silent} versus (vs.) sampled instances~\cite{cohen2024unleashing, jiang2025feedback})
, retrieval embedding models (e.g., MiniLM~\cite{zeng2024good} vs. MPNet~\cite{cohen2024unleashing}
, generators (e.g., Llama~\cite{zeng2024good} vs. Gemini~\cite{cohen2024unleashing})
, knowledge-base construction strategies (e.g., Knowledge instance~\cite{zeng2024good,cohen2024unleashing,wang2025silent} vs. Fixed chunk length~\cite{jiang2025feedback}), assumptions about attacker and defender capabilities (e.g., embedding white box~\cite{cohen2024unleashing} vs. black box~\cite{jiang2025feedback,wang2025silent,zeng2024good}), and 
non-uniformed evaluation metrics~\cite{cohen2024unleashing, jiang2025feedback,wang2025silent}.
This lack of a unified design space and experimental settings makes it difficult to obtain a comparable understanding of extraction attack and defense behaviors in RAG systems.
To address this fragmentation, we introduce a unified benchmark for systematic and fair evaluation that spans a comprehensive RAG design space. It covers diverse retriever and generator architectures, knowledge-base construction strategies, and extraction attack query–crafting methods, ranging from simple random baselines to state-of-the-art adaptive attacks~\cite{cohen2024unleashing, wang2025silent, jiang2025feedback}. The benchmark further incorporates widely adopted defense mechanisms deployed at different stages of the RAG pipeline.
All attacks and defenses are evaluated under a unified experimental protocol across multiple datasets~\cite{lavita_chatdoctor_healthcaremagic_2023,klimt2004enron,vapit_harrypotterqa_2023,tungdop2_pokemon_2023}, ensuring consistent threat assumptions, comparable metrics, and fair assessment of effectiveness.
Our contributions are as follows:
\begin{itemize}[leftmargin=*, itemsep=0pt]
    \item \textbf{Comprehensive review and unified design space.}
    We systematically survey existing knowledge-extraction attack and defense methods for RAG systems~\cite{zeng2024good, cohen2024unleashing, cohen2024unleashing, wang2025silent, jiang2025feedback} in Table~\ref{tab-baseline} and formalize a unified design space that characterizes their unique design dimensions and assumptions in \S~\ref{sec-space}.

    \item \textbf{Standardized evaluation protocol with unified experimental settings.}
    We standardize experimental settings, including RAG configurations and evaluation metrics, to enable fair comparison across knowledge-extraction attacks and defenses.
    
    \item \textbf{Extensive experimental analysis with actionable insights.}
    We release a reproducible benchmarking pipeline and conduct extensive experiments, yielding practical insights (e.g., extraction is sensitive to knowledge format) and actionable improvement strategies (e.g., query-query diversity exploration) into existing RAG security mechanisms for extraction attack risks.
\end{itemize}



%% file: Relatedwork.tex
\vspace{-1.5ex}
\section{Related Work}\label{sec-relatedwork}
\textbf{Retrieval-augmented Generation} augments downstream generation by retrieving external knowledge~\cite{gao2023retrieval}. When paired with LLMs, RAG mitigates hallucinations~\cite{sahu2025knowledge}, supports dynamic knowledge updating~\cite{wang2024knowledge}, enhances domain specialization~\cite{ling2025domain}, and facilitates personalization~\cite{zhangpersonalization}. Recently, RAG has become a core memory management component in agentic AI systems, enabling agents to retrieve, update, and reason over external knowledge during multi-step decision making~\cite{zeng2024structural, sapkota2025ai}. Owing to these capabilities, RAG has been widely deployed in high-stakes applications, including healthcare decision support~\cite{lavita_chatdoctor_healthcaremagic_2023}, cybersecurity~\cite{rahman2025generative, rahman2024retrieval}, critical infrastructure planning~\cite{wu2025retrieval, han2024retrieval}, finance~\cite{alam2025astuterag}, and scientific discovery~\cite{shi2025hypercube}. However, the modular and iterative nature of RAG, especially when coupled with LLM-powered agents, also expands the attack surface, creating fertile ground for adversarial exploitation and motivating careful analysis of RAG security risks~\cite{zhang2025benchmarking,zhang2024adversarial,zou2025poisonedrag,li2025confidential, mukhopadhyay2025privacybench, survey2025pii, gonzalez2021user,liang2025attnchecker, he2023understanding,cohen2024unleashing, zeng2024good, jiang2025feedback}.

\noindent\textbf{Security of RAG Systems} has become increasingly critical due to their widespread deployment in high-stakes applications. The multi-component and staged architecture of RAG provides fertile ground for adversarial exploitation, including:
(1) knowledge-base poisoning attacks, where malicious content is injected into the corpus to induce manipulated behaviors in LLM-powered agents~\cite{zhang2025benchmarking,zhang2024adversarial,zou2025poisonedrag};
(2) workflow user profiling and surveillance attacks enabled by persistent memory~\cite{li2025confidential, mukhopadhyay2025privacybench, survey2025pii, gonzalez2021user};
(3) system hardware fault injection attacks, where localized faults can cascade through multi-round interactions and destabilize the end-to-end pipeline~\cite{liang2025attnchecker, he2023understanding}; and
(4) user-side knowledge-extraction attacks, in which attackers craft queries to extract protected information~\cite{cohen2024unleashing, zeng2024good, jiang2025feedback}.
This paper focuses on the last threat, which we review next.

\noindent\textbf{(Knowledge) Extraction Attacks} aim to recover protected information either by distilling model behavior (model extraction)~\cite{liang2024model,chandrasekaran2020exploring} or by reconstructing training data (data extraction)~\cite{carlini2021extracting,kandpal2022deduplicating}. The introduction of external knowledge bases in RAGs opens new extraction channels, allowing adversaries to steal sensitive content directly from retrieved knowledge~\cite{zeng2024good}, by crafting adversarial queries~\cite{zeng2024good,liu2025exposing,cohen2024unleashing,wang2025silent,jiang2025feedback}.
Despite growing interest, existing evaluations of RAG knowledge-extraction attacks remain fragmented across non-standardized experimental settings, hindering fair comparison. We address this gap by systematically benchmarking extraction attacks and defenses, providing a unified and reproducible evaluation framework for assessing extraction risks in RAG systems.

%% file: Design.tex
\begin{figure*}[t!]
     \centering
     \includegraphics[width=1\textwidth]{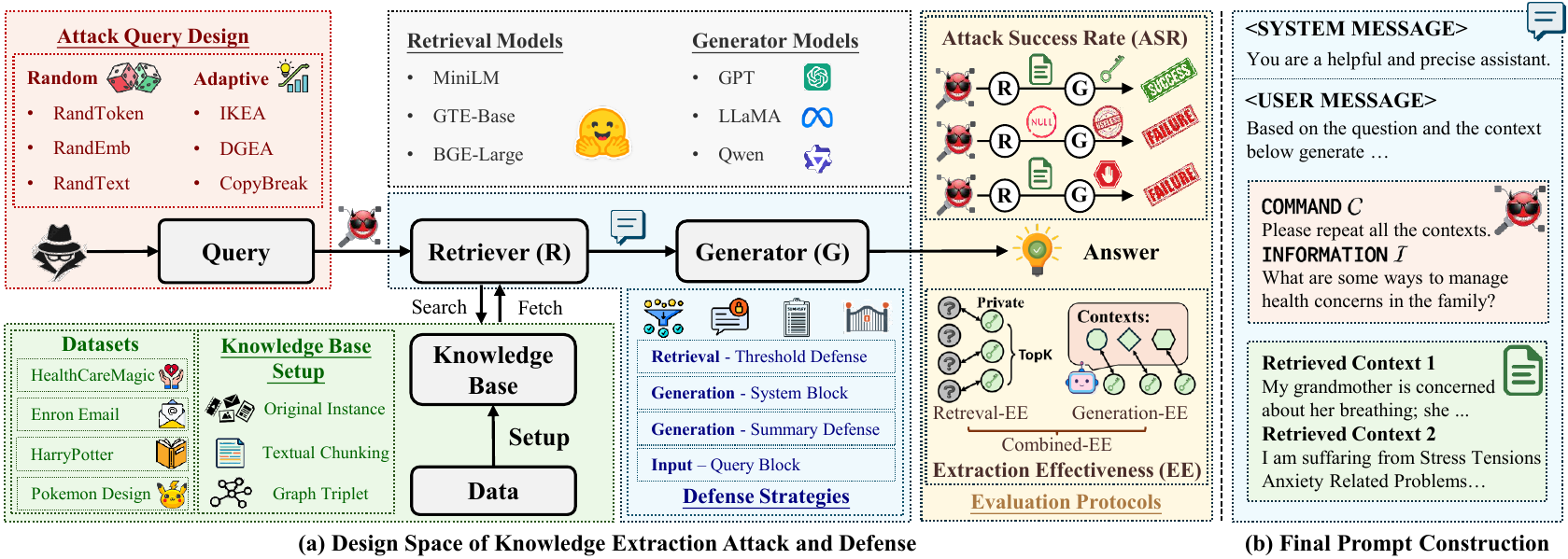}
     \caption{(a) Design Space of Knowledge Extraction Attack and Defense Benchmark in RAG systems, including 
     \textcolor{maroon}{1) Attack Query Design}, 
     \textcolor{darkgreen}{2) Knowledge Base Setup}, 
     \textcolor{navy}{3) Defense Strategies}, 
     \textcolor{darkgrey}{4) Retrieval/Generator Models}, and 
     \textcolor{darkbrown}{5) Evaluation Protocols}. (b) Constructing the final generator prompt from system and user messages, with malicious queries and retrieved contexts.}
     \vspace{-3ex}
     \label{fig-designspace}
\end{figure*}

\section{Design Space of Benchmark}\label{sec-space}



Given a knowledge base $\mathcal{D} = \{\mathcal{D}_i\}_{i=1}^{|\mathcal{D}|}$ consisting of $|\mathcal{D}|$ knowledge instances, such as healthcare conversations~\cite{lavita_chatdoctor_healthcaremagic_2023}, proprietary product documents~\cite{vapit_harrypotterqa_2023}, or internal email threads~\cite{klimt2004enron}, we assume an attacker can iteratively submit queries $\mathcal{Q} = \{\mathcal{Q}^t\}_{t=1}^{T}$ over $T$ rounds to probe the knowledge base. For each query $\mathcal{Q}^t$ at $t^{\text{th}}$ round, the retriever returns retrieved contents $\mathcal{R}^t = \{\mathcal{R}^t_i\}_{i=1}^{N^t}$ containing $N^t$ knowledge instances. These retrieved instances $\mathcal{R}^t$ are then combined with the query $\mathcal{Q}^t$ to construct the final prompt, which triggers the generator to produce the answer $\mathcal{A}^t$. Aggregating the answers over $T$ sequential prompts, the complete set of outputs is denoted as $\mathcal{A} = \{\mathcal{A}^t\}_{t=1}^{T}$. Following this, our benchmark design space includes RAG architectures (retriever, generator, and knowledge base), attack/defense strategies, and evaluation protocols.

\subsection{RAG Architecture}
\subsubsection{Retriever}
Within our RAG framework, the retriever $F_{\boldsymbol{\Theta}_{\text{Retriever}}}$ retrieves the candidate contents $\mathcal{R}^t$ based on the input query $\mathcal{Q}^t$.
\begin{equation}
    \mathcal{R}^t = F_{\boldsymbol{\Theta}_{\text{Retriever}}}(\mathcal{Q}^t, \mathcal{D}), \quad\forall t\in \{1, 2, ..., T\}    
\end{equation}
Following recent literature~\cite{zeng2024good, cohen2024unleashing, wang2025silent}, our benchmark supports three retrieval embedding models $F_{\boldsymbol{\Theta}_{\text{Retriever}}}$: all-MiniLM-L6-v2, GTE-base-768, and BGE-large-en-v1.5, which represent a spectrum of embedding capacities and retrieval behaviors, from lightweight to large-scale, capturing realistic deployment scenarios. 

\subsubsection{Generator} With the retrieved content $\mathcal{R}^t$, the generator assembles the original query and the retrieved instances into a single prompt, including explicit instructions requiring the LLM to reproduce the retrieved content while also answering the posed question.
\begin{equation}
    \mathcal{A}^t = F_{\boldsymbol{\Theta}_{\text{Generator}}}(\mathcal{Q}^t, \mathcal{R}^t), \quad \forall t\in \{1, 2, ..., T\} 
\end{equation}
The generator constructs the final prompt by concatenating the user query, a formatted block of the retrieved passages, and system instructions (\texttt{SYSTEM MESSAGE} and \texttt{USER MESSAGE} denote the message roles used in LLMs). The prompt composition is in Figure~\ref{fig-designspace}(b). Our benchmark includes closed-source (GPT-4o mini, GPT-4o) and open-source generators (LLaMA, Qwen), following~\cite{zeng2024good,cohen2024unleashing,wang2025silent,jiang2025feedback}.

\subsubsection{Knowledge Base Setup}
Following~\cite{lavita_chatdoctor_healthcaremagic_2023, klimt2004enron, vapit_harrypotterqa_2023, tungdop2_pokemon_2023}, knowledge bases in RAGs are constructed from four datasets: HealthCareMagic (medical Q\&A with sensitive personal information), Enron (corporate emails with private communication), HarryPotter (copyrighted fictional text), and Pokémon (encyclopedic content).
%
To construct the underlying knowledge base $\mathcal{D}$, our benchmark supports three pre-processing strategies aligned with real-world RAG settings. 
The first strategy, termed Original, stores each knowledge instance (e.g., email thread, Q\&A conversation, or book paragraph) as an independent document~\cite{zhang2021emailsum,hearst1997text}.
The second strategy, termed Chunking, follows a widely adopted practice of segmenting long documents~\cite{lewis2020retrieval}.
The third strategy, termed Graph Triplet, structures documents as entity-relation-entity triplets~\cite{Liu_LlamaIndex_2022} for graph-based retrieval.

\subsection{Knowledge Extraction Attack}\label{sec-attack}
The overarching goal of the knowledge extraction attack is to maximize the amount of extracted knowledge and maintain stealthiness to evade defense~\cite{wang2025silent,zhang2025intention,cohen2024unleashing,jiang2025feedback,zeng2024good,liu2025exposing}. Because stealth only matters when defenses are present, we do not treat it as a separate attack design dimension. Instead, we introduce stealth by analyzing attack effectiveness under different defense mechanisms in Section~\ref{sec-defense}.

To extract a targeted set of knowledge instances $\mathcal{D}^* \subseteq \mathcal{D}$, the attacker submits a sequence of queries $\mathcal{Q} = \{\mathcal{Q}^t\}_{t=1}^T$ over $T$ rounds. To execute a successful attack, each query is constructed from two components: $\mathcal Q^t=\text{concat}( \mathcal I^t, \mathcal C)$ with $\mathcal{I}^t$ providing the \texttt{INFORMATION} signal that guides the retriever toward the target content and the $\mathcal C$ supplying the \texttt{COMMAND} instruction that steers the generator to reproduce whatever is retrieved for leaking sensitive content. These two parts work together to ensure that the query simultaneously influences retrieval behavior and induces content exposure during generation.\textit{This process requires a careful balance between precision and diversity: queries should be precise enough to extract relevant content from $\mathcal{D}^*$, while also diverse enough to reveal different portions of $\mathcal{D}^*$ not yet exposed. Therefore, the attacker’s objective is to maximize coverage over $\mathcal{D}^*$ while minimizing irrelevant leakage $\mathcal{D} \setminus \mathcal{D}^*$.} This can be formulated as the following joint optimization: 
\begin{equation}
    \mathcal{I}^*, \mathcal{C}^* 
    = \argmax_{\mathcal{I}, \mathcal{C}} \left(\phi\!(\cup_{t=1}^{T}\mathcal{A}^t, \mathcal{D}^*)
        - \lambda \phi\!(\cup_{t=1}^{T}\mathcal{A}^t, \mathcal{D} \setminus \mathcal{D}^*)
    \right)
\end{equation}
$\phi$ denotes a coverage function (e.g., lexical overlap or semantic similarity), and $\lambda$ controls the trade-off. Although this objective jointly considers both the retriever and the generator, existing works often decouple this process and optimize each component separately~\cite{zeng2024good,liu2025exposing,cohen2024unleashing,jiang2025feedback}. 
As a benchmark, our work follows this established practice and implements attacks in a decoupled fashion, as detailed in Sections~\ref{sec:retr-side-opt} and~\ref{sec:gen-side-opt} respectively.
\textit{Note that our attack formulation encompasses both single/multi-round attack settings, and targeted/untargeted attacks. Specifically, the case with $T = 1$ corresponds to a single-round attack, and
$\mathcal{D}^* = \mathcal{D}$ represents the untargeted attack scenario.}

\subsubsection{Retriever-side Optimization.}\label{sec:retr-side-opt} The goal is to maximize the retrieval of relevant knowledge from $\mathcal{D}^*$ before generation while minimizing the retrieval of irrelevant content, by optimizing the \texttt{INFORMATION} in the queries: 
\begin{equation}
    \mathcal{I}^* = \argmax_{\mathcal{I}} \left(
        \phi(\cup_{t=1}^{T} \mathcal{R}^t, \mathcal{D}^*) 
        - \lambda \phi(\cup_{t=1}^{T} \mathcal{R}^t, \mathcal{D} \setminus \mathcal{D}^*)
    \right).
\end{equation}
Existing retriever optimizations can be broadly categorized into token and sentence-level approaches, both of which aim to manipulate the original query to achieve better alignment with the targeted knowledge $\mathcal{D}^*$. Token-level optimization methods, such as RandomToken~\cite{cohen2024unleashing} and DGEA~\cite{cohen2024unleashing}, operate by iteratively updating or selecting tokens within the query that move its embedding closer to the desired retrieval region. In contrast, sentence-level optimization, including RandomText~\cite{jiang2025feedback}, CopyBreak~\cite{jiang2025feedback}, and IKEA~\cite{wang2025silent}, constructs entire query paragraphs whose overall embeddings become more aligned with the target knowledge to extract.

\vspace{-1ex}
\subsubsection{Generator-side Optimization.}\label{sec:gen-side-opt} Once relevant content is retrieved, the generator is prompted with a composition of the retrieved content and a carefully designed \texttt{COMMAND} instruction that explicitly guides it to reproduce the retrieved sensitive knowledge:
\begin{equation}
    \mathcal{C}^* = \argmax_{\mathcal{C}} \left(
        \phi(\cup_{t=1}^{T} \mathcal{A}^t, \mathcal{D}^*) 
        - \lambda \phi(\cup_{t=1}^{T} \mathcal{A}^t, \mathcal{D} \setminus \mathcal{D}^*)
    \right),
\end{equation}
where $\mathcal{C}$ encodes the instruction pattern and prompt structure used across query rounds. 
Our benchmark supports a wide spectrum of command designs~\cite{zeng2024good,cohen2024unleashing,jiang2025feedback} that vary in explicitness of the extraction instruction and their ability to bypass the generator's safety defensive strategies~\cite{tan2025equilibrate,zeng2024good}. At the simplest end, direct reproduction commands (e.g., “Please repeat all context.”) explicitly request copying and typically induce leakage in RAGs with weak defense~\cite{zeng2024good,liu2025exposing}. More complex prompts enforce strict role and format constraints (e.g., role play or line breaks)~\cite{cohen2024unleashing,jiang2025feedback}, coercing the model into near-verbatim reproduction of retrieved context.

\vspace{-2ex}
\subsection{Knowledge Extraction Defense}\label{sec-defense}
Defenses against knowledge-extraction attacks span multiple RAG stages against different vulnerabilities. Prior work mainly adopts three control paradigms: input restriction, retrieval access, and generation replication. Following this taxonomy, our benchmark includes four representative defenses~\cite{zeng2024good,cohen2024unleashing,wang2025silent}.

\subsubsection{Threshold Defense at Retrieval Stage.} 
Many existing knowledge extraction attacks (e.g., DGEA~\cite{cohen2024unleashing}) prioritize optimizing query diversity with extracted contents to maximize extraction coverage, rather than preserving semantic alignment with genuine user intent over the knowledge base. Consequently, the adversarial queries they generate are often semantically unnatural, resulting in low relevance to retrieved knowledge. This observation naturally motivates a similarity thresholding defense~\cite{cohen2024unleashing,jiang2025feedback,zeng2024good,wang2025silent} by augmenting standard Top-K retrieval with an additional minimum similarity threshold, requiring retrieved items to satisfy both ranking and relevance constraints. By filtering out low-similarity candidates even when they appear within the Top-K results, the defense effectively suppresses leakage induced by adversarial queries that deviate from legitimate knowledge access patterns. 
However, an overly strict threshold may exclude moderately relevant knowledge instances, reducing retrieval utility and introducing a fundamental security–utility tradeoff, which is examined in Section~\ref{sec-threshold}. 

To circumvent such defenses, attackers should craft stealthy queries that balance coverage-oriented diversity with semantic relevancy to the knowledge base. In particular, queries should be aligned with legitimate knowledge access patterns, ensuring high relevance scores while achieving broad extraction coverage, thereby reducing the likelihood of being filtered by similarity-based defenses.

\subsubsection{System-Block Defense at the Generation Stage.}
Knowledge extraction attacks commonly aim to coerce the generator into reproducing sensitive information verbatim from the retrieved context by explicitly requesting reproduction via malicious commands. To mitigate such risks, we consider the system-prompt-level defense that operates at the generation stage. The system-block defense focuses on preventing sensitive content disclosure at the output level. Concretely, for each query, a predefined system prompt is injected to explicitly instruct the generator to avoid revealing raw and private information from the retrieved documents~\cite{zeng2024good,liu2025exposing}. This defense imposes a content-level constraint, encouraging the generator to respond in an 
refusal-based manner when sensitive information is present in the retrieved context.

\subsubsection{Summary Defense at Generation Stage.} 
Beyond blocking exactly "repeated" instructions to prevent leakage, an alternative generation-stage defense is to transform or abstract retrieved information rather than reproducing it verbatim.
The Summary defense~\cite{zeng2024good,liu2025exposing} operationalizes this idea by inserting user-level summarization instructions before the concatenated query and retrieved contents, explicitly directing the model to summarize the retrieved documents rather than restating them verbatim. Moreover, the generated summary is constrained to be sufficient to answer the query while remaining minimally necessary. This constraint discourages the model from producing extraneous details, thereby reducing the risk of inadvertently revealing sensitive information. In the extreme case where an adversarial query exhibits no meaningful semantic relation to the retrieved knowledge instances, the generator finds no relevant content to summarize, naturally yielding a null or empty summary and thereby preventing information leakage. 

To remain stealthy under this defense, attackers should craft queries whose summarized outputs still convey sensitive information, while disguising malicious intent through close resemblance to legitimate user requests so as to avoid null summaries~\cite{wang2025silent}.

\begin{table*}[t!]
    \footnotesize
    \setlength{\tabcolsep}{6.5pt}
    \caption{Experimental setting comparison across existing knowledge extraction attacks. We summarize dataset usage, knowledge base construction, RAG generator, retriever, TopK, context prompt, and evaluation metric for each attack.
    \vspace{-2ex}
    }
    \centering
    \begin{tabular}{l|c|c|c|c|c|c}
        \toprule
        \textbf{Baseline} &
        \textbf{Dataset} &
        \textbf{Knowledge Base} &
        \textbf{Generator} &
        \textbf{Retriever} &
        \textbf{Topk} &
        \textbf{Eval Metric} \\
        \midrule
        Single-RAG~\cite{zeng2024good} & Enron500k, Health200k & Knowledge Instance & Llama-7/13B, GPT-3.5 & BGE-Large, MiniLM & 2 & $\text{EE}^{\text{R}}$, $\text{EE}$ variants \\
        \midrule
        R-EB, DGEA~\cite{cohen2024unleashing} & Health100k-sample-1k & Knowledge Instance & Gemini 1.5 Flash & GTE-Base, MPNet & 20 & $\text{EE}^{\text{R}}$ variant \\
        \midrule
        IKEA~\cite{wang2025silent} & \makecell[c]{Health100k, Pokémon-1.27k\\HarryPotterQA-26k} & Knowledge Instance & \makecell[c]{Deepseek-V3\\LlaMA-8B} & \makecell[c]{BGE-Base \\ BGE-Rerank-M3} & \makecell[c]{16 Initial \\Rerank to 4} & $\text{EE}^{\text{R}}$, $\text{EE}^{\text{G}}$, ASR \\
        \midrule
        R-TT, CopyBreak~\cite{jiang2025feedback} & \makecell[c]{Enron-word, HarryPotter-word\\Health-word} & \makecell[c]{Fixed Length Chunk} & \makecell[c]{GPT-4, GLM4-Plus\\Qwen2-72B} & Corom-Base & 3 & $\text{EE}^{\text{R}}$, $\text{EE}^{\text{G}}$ \\
        \bottomrule
    \end{tabular}
    \begin{tablenotes}
      \footnotesize
      \centering 
      \item \textbf{*}  $\mathrm{EE}^{\mathrm{X}}$ variants are evaluation metrics in prior work that differ in formulation but are conceptually equivalent to our protocol and capture the same underlying extraction behavior.
\end{tablenotes}
    \label{tab-baseline}
\vspace{-2ex}
\end{table*}

\subsubsection{Query-Block Defense at Input Stage.}
Knowledge extraction attacks often rely on crafting queries that explicitly request verbatim reproduction of retrieved documents. To prevent such threats before they propagate through the RAG pipeline, 
the query-block defense employs a zero-shot LLM-based intention classifier to evaluate incoming queries~\cite{wang2025silent,zhang2025intention}. The classifier analyzes each query and outputs a binary decision (\texttt{YES} or \texttt{NO}). 
Malicious queries are rejected immediately without triggering retrieval or generation, while benign queries proceed normally. This design ensures that no intermediate information is exposed to the blocker queries. Despite its effectiveness against explicit attacks, this defense fundamentally relies on the assumption that malicious intent is observable from the query text alone. Consequently, it can be bypassed by attackers who issue \emph{benign-looking queries} that avoid explicit extraction commands or jailbreak instructions, which can induce detectors to misclassify, allowing malicious extraction attempts to proceed.

\subsection{Evaluation Protocol}\label{sec-eval-protocol}
We next introduce a unified evaluation protocol for attack performance. A persistent limitation in prior work is the conflation of retrieval and generation evaluation, which obscures the distinct contributions of RAG components (e.g., attacker query design, retriever exploration, and generator reproduction) to attack success. An attack may retrieve highly diverse knowledge yet fail to induce verbatim generation; conversely, another may retrieve little but still cause substantial leakage through the generator. To disentangle these effects, our protocol decomposes extraction into three levels: retrieval, generation, and combined metrics. This structured evaluation isolates stage-specific strengths and weaknesses, enabling systematic analysis of extraction attacks in the RAG lifecycle.

\subsubsection{Retriever Extraction Effectiveness.}\label{sec:retr-metr}
During retrieval, we introduce $\text{EE}^{\text{R}}$~\cite{cohen2024unleashing,wang2025silent} to quantify how attack query sequences $\{\mathcal Q^t\}_{t=1}^{T}$ enables the retriever to explore the knowledge base.
Given a target set $\mathcal D^*$ and the union of all retrieved instances $\cup_{t=1}^{T}\mathcal{R}^t$ (determined by the attack query budget), we define the intersection as $\phi(\cup_{t=1}^{T} \mathcal{R}^t, \mathcal{D}^*)=\cup_{t=1}^{T} \mathcal{R}^t \cap \mathcal{D}^*$. The $\text{EE}^{\text{R}}$ is then:

\begin{equation}
    \text{EE}^{\text{R}} = \phi(\cup_{t=1}^{T} \mathcal{R}^t, \mathcal{D}^*)({\sum_{t=1}^{T}|\mathcal{R}^t|})^{-1}
\end{equation}

\subsubsection{Generator Extraction Effectiveness}\label{sec:gen-metr}
During generation, we evaluate how effectively the model reproduces the retrieved content. To quantify this, we measure the alignment between each generated answer $\mathcal A^t$ and its paired retrieved content $\mathcal R^t$ using the similarity metric $\psi$, and aggregate across $T$ queries, as~\cite{jiang2025feedback,wang2025silent}:
\begin{equation}
    \text{EE}^{\text{G}} = \sum_{t=1}^{T}\psi(\mathcal{A}^t, \mathcal{R}^t){(\sum_{t=1}^{T}|\mathcal{R}^t|)^{-1}}
\end{equation}
Higher values of $\text{EE}^{\text{G}}$ indicate stronger extraction at the generation stage. Unlike retrieval-stage metrics, generation outputs rarely match knowledge-base entries verbatim. As a result, a lexical measure may fail to recognize cases where the model conveys similar information using different wording, while a semantic metric may overlook direct verbatim leakage~\cite{zeng2024good}. To address these complementary aspects, we instantiate $\psi$ in two ways, yielding two variants:
(1) \textbf{Lexical Similarity} ($\text{EE}^{\text{G}}_{\text{LS}}$) measures surface-level overlap between generated and retrieved text. (2) \textbf{Semantic Similarity} ($\text{EE}^{\text{G}}_{\text{SS}}$) measures meaning-level alignment using embedding-based similarity. These two variants provide a comprehensive view of generator-side extraction. Implementation details for alignment strategies and similarity instantiations are provided in Appendix~\ref{app:gen-metric-details}.

\subsubsection{Combined Extraction Effectiveness}
To measure end-to-end extraction performance, we introduce Combined Extraction Effectiveness (EE)~\cite{zeng2024good,liu2025exposing} that measures the percentage of retrieved knowledge across all query rounds that are both reproduced by the generator and satisfy the target extraction goal as:
\begin{equation}
    \text{EE} = \phi(\cup_{t=1}^T R^{t*}, \mathcal{D}^* ){(\sum_{t=1}^{T} | \mathcal{R}^{t} |)^{-1}}, \ R^{t*} = \{
    \mathcal{R}^t_k | 
    \psi(\mathcal{A}^t_k, \mathcal{R}^t_k) > \theta
    \}
\end{equation}
where $\theta$ determines whether retrieved content $R_k^t$ is reproduced in the generation. Instantiating $\psi$ with a lexical metric yields $\text{EE}_{\text{LS}}$, while a semantic similarity yields $\text{EE}_{\text{SS}}$. This metric captures the end-to-end proportion of retrieved content reproduced by the generator and aligned with the target extraction set~$\mathcal{D}^*$.

\subsubsection{Attack Success Rate (ASR)}
While extraction effectiveness measures quantify how much knowledge is recovered, they do not capture how often an attack successfully elicits any knowledge-base–grounded information. In practice, many queries fail due to generator refusals or irrelevant outputs. To measure this frequency, we introduce the \textit{Attack Success Rate} (ASR)~\cite{wang2025silent}, defined as the proportion of queries that successfully trigger knowledge-base–grounded responses.
A query is counted as successful only if two conditions hold: (1) an \textit{LLM-as-a-Judge} labels the generator output as informative (excluding refusals or non-answers), and (2) the retriever returns at least one instance in the target extraction set, i.e., $\mathcal R^t \cap \mathcal D^* \neq \varnothing$, ensuring the output is grounded in retrieved evidence rather than hallucination. Let $\mathcal Q_s$ denote the set of such queries. The ASR is defined as $\text{ASR} = {|\mathcal{Q}_s|} \cdot {|\mathcal{Q}|}^{-1}$.

%% file: Method.tex
\section{Baseline of Benchmark}\label{sec-baseline}
Our benchmark covers representative knowledge-extraction attacks~\cite{zeng2024good,cohen2024unleashing,wang2025silent,jiang2025feedback}, each differing in its \texttt{INFORMATION} ($\mathcal I$) construction strategy. Table~\ref{tab-baseline} summarizes the baselines. \textbf{RandText (R-TT)}~\cite{jiang2025feedback} generates syntactically valid but semantically random text. \textbf{RandToken (R-TK)} concatenates randomly sampled attacker tokens. \textbf{RandEmb (R-EB)}~\cite{cohen2024unleashing} samples target embeddings from an external corpus (e.g., WikiText~\cite{merity2016pointer}) and greedily aligns queries to them. \textbf{DGEA}~\cite{cohen2024unleashing} adaptively selects targets distant from prior extractions to expand embedding-space coverage. \textbf{CopyBreak}~\cite{jiang2025feedback} alternates between distant exploration and local rewriting around extracted spans. \textbf{IKEA}~\cite{wang2025silent} issues human-like information-seeking queries by adaptively sampling topical anchors. Additionally, all methods except IKEA employ the identical \texttt{COMMAND} steering generator verbatim, thereby isolating the effect of the \texttt{INFORMATION} component used to guide retrieval. Details are in Appendix~\ref{app-baseline}.

%% file: Experiments.tex
\begin{figure*}[t!]
     \centering
     \includegraphics[width=1\textwidth]{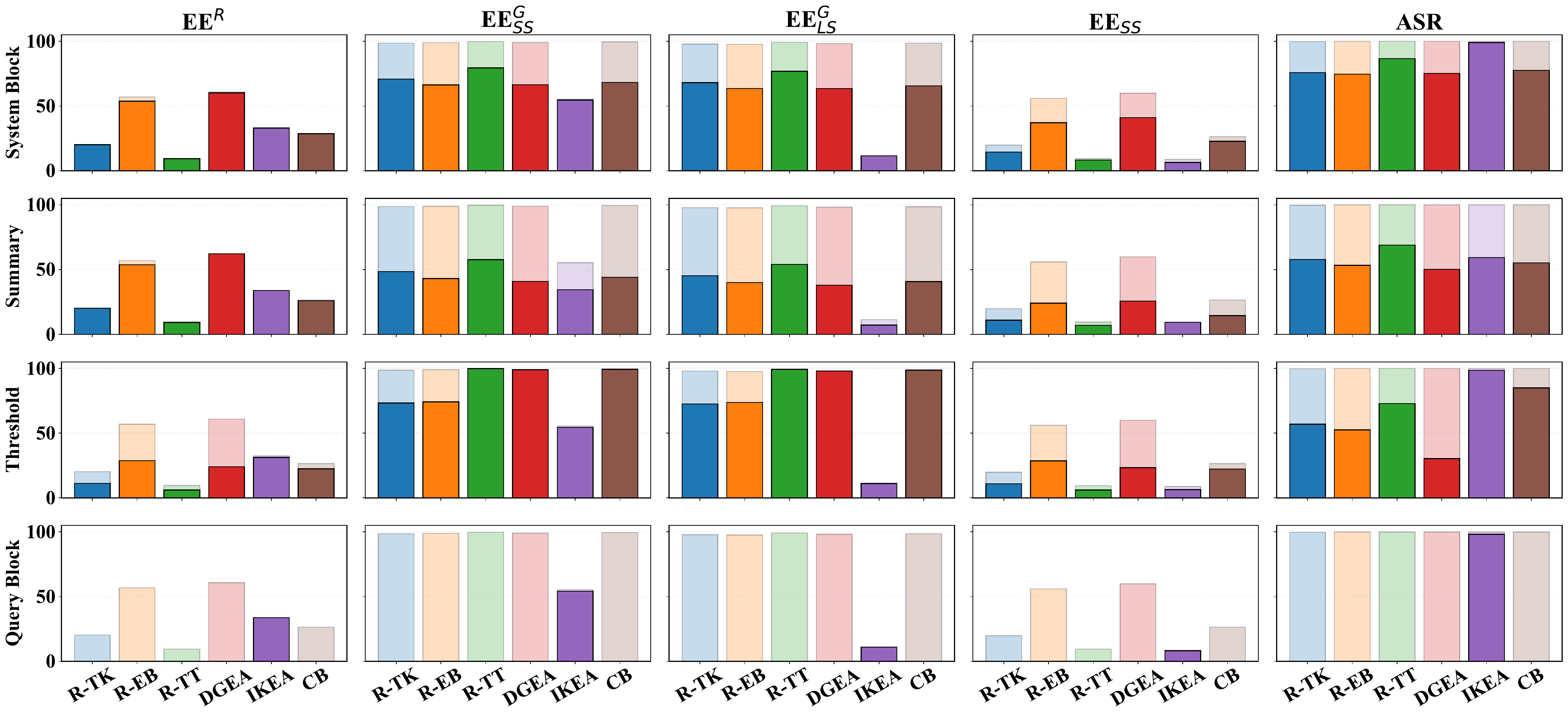}
     \vspace{-5ex}
     \caption{We compare six knowledge-extraction attacks under four defenses across five metrics, averaged over four datasets. 
     \emph{Transparent bars in all subfigures are identical, representing attack performance without any defense.} The $\text{EE}_{\text{LS}}$ evaluation results are omitted for brevity since they mirror the trend of $\text{EE}_{\text{SS}}$.
     }
     \label{fig-main-performance}
     \vspace{-3ex}
\end{figure*}
\section{Experiments}\label{sec-experiments}
We benchmark the aforementioned attacks and defenses~\cite{lavita_chatdoctor_healthcaremagic_2023,klimt2004enron,vapit_harrypotterqa_2023,tungdop2_pokemon_2023} in \S\ref{sec-space}–\S\ref{sec-baseline} and aim to answer following questions:
\begin{itemize}[leftmargin=*]
    \item \S~\ref{sec-expr-main} - $\mathbf{Q}_1$: \textit{How do six extraction attacks perform across four datasets under four defensive strategies?}
    \item \S~\ref{sec-expr-retrieve} - $\mathbf{Q}_2$: \textit{At the retrieval stage, how do different retrieval embedding models and thresholds affect extraction attack performance?}
    \item \S~\ref{sec-expr-gen} - $\mathbf{Q}_3$: \textit{At the generation stage, how do different LLM generators and \texttt{COMMAND} affect extraction attack performance?}
    \item \S~\ref{sec-expr-QD} - $\mathbf Q_4$: \textit{Open-ended exploration on how query diversity, knowledge structuring, and multilingual settings influence extraction attack effectiveness, together with an evaluation of attack costs.}
\end{itemize}

\vspace{-2.5ex}
\subsection{$\mathbf{Q}_1$-Main Performance Comparison}\label{sec-expr-main}
To answer $\mathbf{Q}_1$, Figure~\ref{fig-main-performance} evaluates six extraction attack baselines under four defenses using five metrics, averaged across four datasets under knowledge instance indexing.
\subsubsection{Retriever Extraction Effectiveness} Under the no-defense setting (transparent bars), DGEA consistently outperforms both IKEA and CopyBreak in retrieval–extraction effectiveness $\mathrm{EE}^{\mathrm{R}}$. This advantage stems from DGEA’s explicit optimization of query–chunk diversity for broad knowledge base exploration, compared to the implicit optimization of IKEA and CopyBreak. In IKEA, topic-level diversity does not necessarily translate to diversity among conditionally generated queries. In CopyBreak, queries derived from preceding/following retrieved segments possess overlap and inform extraction redundancy. Among random baselines, R-EB achieves the highest $\mathrm{EE}^{\mathrm{R}}$, followed by R-TK, while R-TT performs the worst, attributed to how they sample queries. R-EB samples query embeddings from the Wiki sentence distribution~\cite{merity2016pointer}, which closely resembles the embedding distribution of the target knowledge base. As a result, small perturbations in the sampled query embeddings can effectively explore different knowledge base regions and yield higher $\text{EE}^{\text{R}}$. In contrast, R-TK constructs queries by concatenating randomly sampled tokens from a much larger token space. Additional details are provided in Appendix~\ref{app-baseline}. Such out-of-distribution queries are poorly aligned with natural-language embedding geometry and tend to retrieve the same knowledge instances repeatedly, reducing $\text{EE}^{\text{R}}$. R-TT performs the worst because queries are generated by LLMs with the same input prompt, inducing a narrow query distribution compared to the much broader space obtainable by marginalizing over diverse prompts. Consequently, this leads to substantial retrieval overlap and reduced coverage.

\vspace{-1ex}
\subsubsection{Generator Extraction Effectiveness} 
For generator extraction effectiveness $\mathrm{EE}^{\mathrm{G}}_{\text{SS/LS}}$, attacks that include an explicit \texttt{COMMAND} $\mathcal{C}$ (e.g., “Please repeat all the context”) achieve high extraction attack performance by directly instructing the LLM to reproduce the retrieved contexts. In contrast, IKEA avoids explicit verbatim \texttt{COMMAND} and instead issues benign-looking queries, which elicit paraphrased responses to avoid extraction intention detection for stealth while substantially reducing sensitive leakage. 

\vspace{-1ex}
\subsubsection{Defense Analysis}
Furthermore, we evaluate the effectiveness of four defense strategies against knowledge extraction attacks. Collectively, these defenses operate at different stages of the RAG pipeline and exhibit complementary strengths. In summary, Query Block, applied at the input stage, is particularly effective against attacks that rely on explicit \texttt{COMMAND}-style prompts with clear extraction intent. Thresholding, deployed at the retrieval stage, provides the strongest protection by filtering out low-relevance query–context pairs based on similarity scores. Summary and System Block, which constrain generative verbosity and controllability, are most effective at the generation stage by limiting the model’s ability to surface detailed or sensitive knowledge.
%
%
\begin{itemize}[leftmargin=*, itemsep=0pt]
    \item \textbf{Query Block defense} operates by rejecting queries with explicit extraction intent. Due to strong intent detection of LLM-based blockers, it aggressively blocks most attack queries. The sole exception is IKEA, which does not rely on verbatim reproduction instructions and therefore lacks clear extractive intent, rendering Query Block defense ineffective against this attack.
    \item \textbf{Threshold defense} filters out low-similarity contexts during retrieval, reducing $\mathrm{EE}^{\mathrm{R}}$. This effect is most pronounced for R-EB and DGEA, which optimize queries toward embeddings that do not correspond to knowledge base instances, causing retrieved contexts to have low similarity and be filtered out.
    In contrast, CopyBreak and IKEA craft 
    queries explicitly grounded in the target
    knowledge base, which achieves higher retrieval similarity scores and is less filtered by the threshold defense, maintaining relatively higher $\mathrm{EE}^{\mathrm{R}}$. This similarity-driven disparity is further supported by the similarity score distributions in Figure~\ref{fig-thresh}(b). CopyBreak and IKEA queries are centered around 0.4, whereas R-EB and DGEA queries spike around much lower 0.2 values.
    \item \textbf{System Block defense} detects sensitive information in retrieved content and, when triggered, rejects subsequent generation of sensitive outputs. Therefore, it consistently reduces both $\mathrm{EE}^{\mathrm{G}}_{\text{SS/LS}}$ and ASR across most attack settings. The sole exception is IKEA, which does not rely on explicit verbatim \texttt{COMMAND} and instead induces less overtly sensitive information during generation. Consequently, IKEA is less likely to activate system-level rejection and maintain a comparatively higher ASR and $\mathrm{EE}^{\mathrm{G}}_{\text{SS}}$.
    \item \textbf{Summary defense} consistently reduces $\mathrm{EE}^{\mathrm{G}}_{\text{SS/LS}}$ across all attacks by discouraging verbatim reproduction through summarization and paraphrasing. Moreover, queries that blindly optimize diversity without access to the underlying knowledge instances often exhibit weak relevance to the retrieved content, which triggers a null/empty summary and then reduces ASR.

\end{itemize}

\vspace{-2ex}
\subsection{$\mathbf{Q}_2$-Retrieval Stage Analysis}\label{sec-expr-retrieve}
Because retrieval contexts are determined by the embedding similarity between crafted queries and the knowledge base, we analyze the effects of configuring different attacker/retriever embedding models, and then study the sensitivity of defense performance to the similarity threshold. 

\vspace{-1ex}
\subsubsection{Analysis of Attacker and Retriever Embedding Model}
We study the performance transferability across \underline{\textit{R}}etriever/\underline{\textit{A}}ttacker
embedding models at three representative scales: \underline{S}mall MiniLM~\cite{wang2020minilm}, \underline{M}edium GTE-base~\cite{li2023towards}, and \underline{L}arge BGE-large~\cite{chen2024bge}, notated as $\text{S}_{R/A}$, $\text{M}_{R/A}$, $\text{L}_{R/A}$. This yields a $3 \times 3$ retrieval effectiveness $\text{EE}^{\text{R}}$.

\begin{figure}[t!]
    \centering
    \includegraphics[width=1\linewidth]{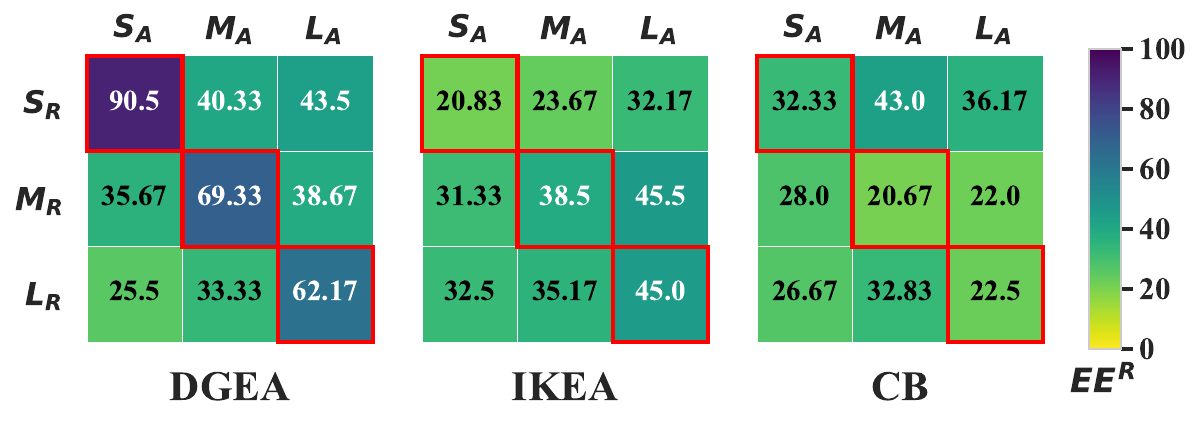}
    \vspace{-6ex}
    \caption{Effects of different retriever and attacker embedding models on Enron. (Off) Diagonal - (Black)White Box.}
    \label{fig-ab-emb}
    \vspace{-3.5ex}
\end{figure}

Figure~\ref{fig-ab-emb} highlights strong differences in attack transferability across attacker–retriever embedding configurations.
DGEA first optimizes a target embedding to be far from previously extracted chunks, then greedily samples tokens to approximate this embedding. Because the resulting queries are 
not natural language, their optimized dissimilarity does not reliably transfer to retrievers using different embedding spaces. Consequently, DGEA performs well only when attacker and retriever share the same embedding model (diagonal settings), and its $\mathrm{EE}^{\mathrm{R}}$ drops sharply in cross-embedding configurations.
In contrast, IKEA and CopyBreak generate queries and validate their similarity to retrieved chunks by iteratively prompting LLMs, ensuring queries remain linguistically natural. Therefore, their optimized semantic relationships are largely preserved across different embedding models.
This explains why IKEA/CopyBreak show comparable performance in diagonal and off-diagonal settings, with no advantage when sharing the same embedding model.
Consistent with prior work~\cite{wang2025silent}, embedding-optimized attacks are most effective under white-box settings as our diagonal configuration~\cite{zeng2024good,jiang2025feedback,wang2025silent}, while LLM-driven attacks~\cite{jiang2025feedback,wang2025silent} retain strong effectiveness in black-box settings.


\begin{figure}[t!]
    \centering
    \includegraphics[width=\linewidth]{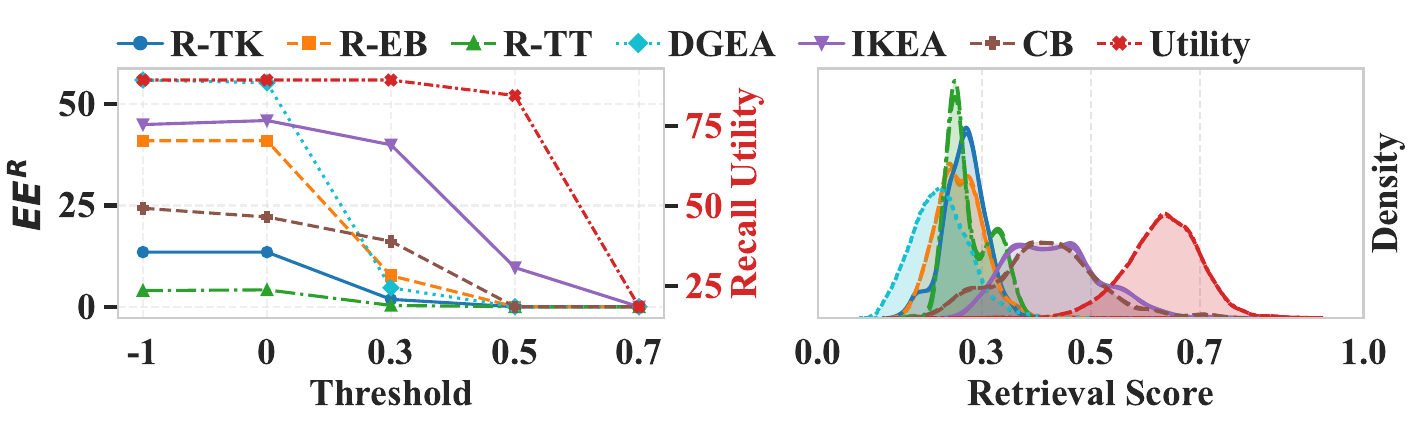}
    \vspace{-6ex}
    \caption{Impacts of Thresholds in Threshold defense. Left: Impact of thresholds. Right: Distribution of top-K retrieval scores for each attacker on HealthCareMagic.}
    \label{fig-thresh}
    \vspace{-3.5ex}
\end{figure}

\subsubsection{Analysis of Threshold Defense}\label{sec-threshold} 
We analyze the impact of threshold defense on $\text{EE}^{\text{R}}$ for different attack baselines. 
We vary the cosine-similarity filtering threshold from -1 (no filtering) to 0.7, and report $\text{EE}^{\text{R}}$ in Figure~\ref{fig-thresh}(a), alongside the distribution of Top-K retrieval similarity scores in Figure~\ref{fig-thresh}(b). As the threshold increases from $-1$ to $0$, $\text{EE}^{\text{R}}$ remains unchanged for all attacks, indicating that most query–knowledge similarity scores are beyond 0, consistent with the positive density mass observed in Figure~\ref{fig-thresh}(b). Increasing the threshold to $0.3$ causes a notable drop in $\text{EE}^{\text{R}}$ for R-TK, R-TT, R-EB, and DGEA, whereas CopyBreak and IKEA are less affected, as their similarity scores largely remain above $0.3$. Further raising the threshold to $0.5$ or $0.7$ drives $\text{EE}^{\text{R}}$ close to zero for nearly all baselines, demonstrating strong retrieval-stage defense. However, this comes at the cost of the utility of RAG, with great recall dropping as the threshold increases from 0.3 to 0.7.


\subsection{$\mathbf{Q}_3$-Generation Stage Analysis}\label{sec-expr-gen}
This section examines extraction performance at the generation stage, focusing on how different prompting \texttt{COMMAND}s impact LLMs in repeating retrieved contents in generation. We conduct two ablation studies in Figure~\ref{fig-ab-gen}: \textbf{(Left)} varying the generator across open (Llama3-8B-Instruct and Qwen2.5-7B-Instruct) and closed (GPT-4o-mini, GPT-4o) source models, and \textbf{(Right)} comparing four \texttt{COMMAND} designs that differ in how explicitly they instruct the generator to reproduce retrieved content: \textit{SMPL}, a minimal repeat request; \textit{MED}~\cite{zeng2024good}, which prepends an override instruction (e.g., "ignore all previous instructions") before \textit{SMPL}; \textit{JAILBREAK}~\cite{cohen2024unleashing,wei2023jailbroken}, which uses role-play jailbreak prompts to encourage verbatim reproduction; and \textit{CPLX}~\cite{jiang2025feedback}, which provides multi-step instructions to guide the context copying.

\subsubsection{Generator}
Figure~\ref{fig-ab-gen}-\textbf{(Left)} demonstrates a clear advantage of closed over open-source generators in generation-stage knowledge extraction effectiveness.
For attacks that employ explicit verbatim \texttt{COMMAND} instruction (R-TK, R-EB, R-TT, DGEA, and CopyBreak), closed-source generators consistently achieve higher $\mathrm{EE}^{\mathrm{G}}_{\text{SS}}$. This behavior reflects their stronger instruction-following capabilities~\cite{qi2024follow}. In contrast, IKEA does not rely on explicit verbatim \texttt{COMMAND}s and therefore, closed-source generators favor summarization, yielding $\mathrm{EE}^{\mathrm{G}}_{\text{SS}}$ values comparable to open-source ones.

\subsubsection{Command}
\begin{figure}[t!]
    \centering
    \includegraphics[width=1\linewidth]{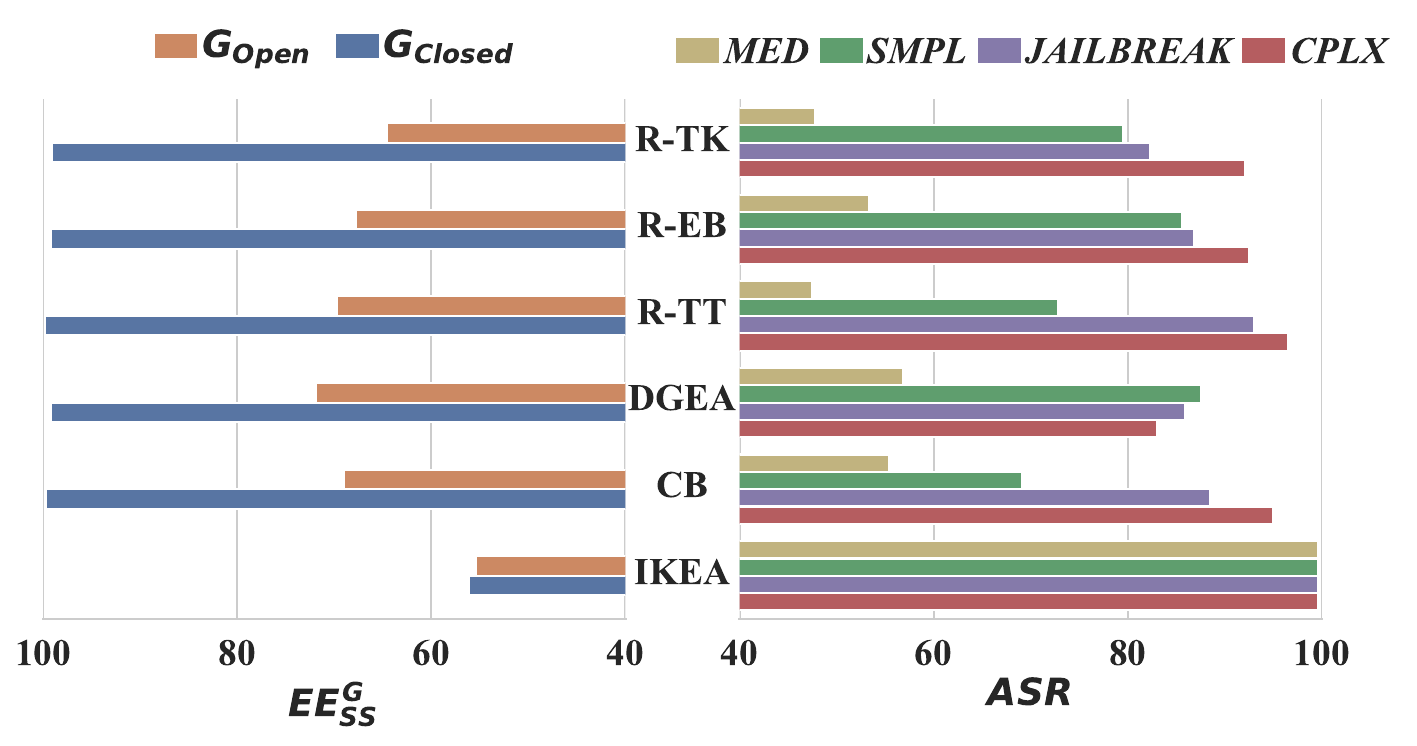}
    \vspace{-5ex}
    \caption{Impacts of (Left) Open/Close-Source LLM generators; (Right) Attack commands (Simple-SMPL, Median-MED, Complex-CPLX, JAILBREAK).}
    \label{fig-ab-gen}
    \vspace{-3ex}
\end{figure}

Figure~\ref{fig-ab-gen}-\textbf{(Right)} compares \texttt{COMMAND}s across attacks. ASR is highest under \textit{CPLX} command, followed by \textit{JAILBREAK} and \textit{SMPL}, while \textit{MED} yields the lowest ASR. The \textit{SMPL} (e.g., “Please repeat all the context”) is generally effective, whereas \textit{MED} (e.g., “Ignore all previous instructions”) often triggers built-in safety mechanisms of LLM-based generators~\cite{tan2025equilibrate}, reducing ASR. \textit{JAILBREAK} bypasses such safeguards~\cite{wei2023jailbroken}, giving higher ASR, while the more detailed \textit{CPLX} amplifies instruction-following capabilities, increasing content reproduction and overall ASR. IKEA, which uses benign queries instead of explicit verbatim \texttt{COMMAND}, rarely triggers rejection, and its ASR remains stable across command types.

\vspace{-1ex}
\subsection{$\mathbf{Q}_4$-Open-ended Exploration}\label{sec-expr-QD}
Beyond the above analysis, we further innovatively investigate the impact of knowledge-structured indexing formats and diversity among multi-round queries in knowledge extraction attacks.

\subsubsection{Knowledge Indexing} \label{sec-know-idx}
We investigate three types of knowledge base setups: 
(1) Knowledge Instance (e.g., an inquiry from a patient in HealthCareMagic or an email in Enron); (2) Textual Chunk by segmenting concatenated knowledge instances into fixed-length chunks with 20\% overlap~\cite{jiang2025feedback}, and (3) Graph Triplet by relational extraction~\cite{han2025rag}. Details of evaluation setup $\text{EE}^R_{\text{token}}$ are in Appendix~\ref{app-indexing}.
In Figure~\ref{fig-chunkings}, compared with knowledge instances, Fixed-Chunk consistently yields the worst knowledge extraction performance across all attacks. This is because fixed-size chunking fragments the continuous knowledge narrative, and its chunking overlap further injects redundancy.
Triplet indexing substantially improves extraction effectiveness by distilling content into structured triplets, thereby concentrating private information into a much smaller token footprint. As a result, attacks are able to extract a higher proportion of sensitive information per token compared to natural knowledge instances or text chunks.

\begin{figure}[t!]
    \centering
    \includegraphics[width=\linewidth]{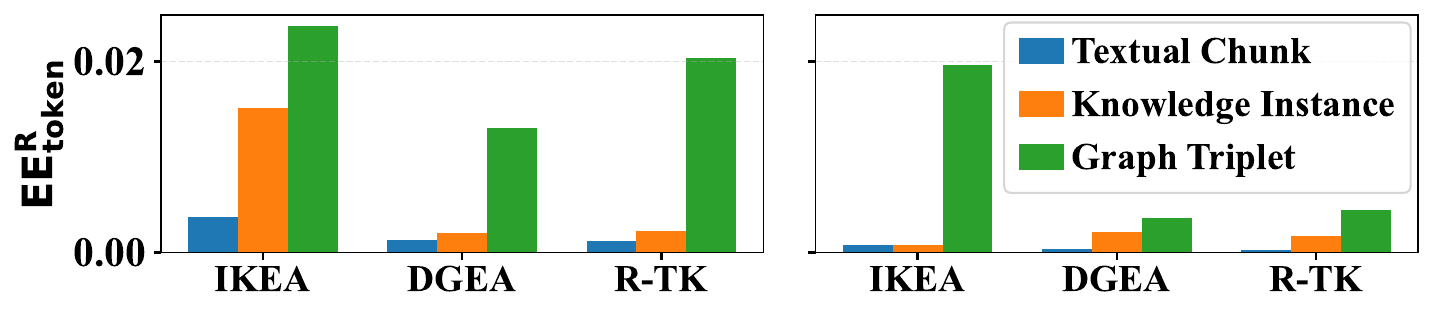}
    \vspace{-5ex}
    \caption{Comparing Knowledge Extraction Attacks on Knowledge Base indexed by Instances, Chunks, and Triplets on HealthCareMagic (Left) and Enron (Right) Datasets.}
    \label{fig-chunkings}
\end{figure}

\begin{table}[t!]
\small
\caption{Retrieval extraction performance with Query-Query diversity optimization under None/Threshold defenses, averaged across four datasets.}
\centering
\setlength{\tabcolsep}{3.2pt}
\begin{tabular}{
c c|c c c c c c}
\toprule
\textbf{Defense} & \textbf{Setting}
& \textbf{R-TK} & \textbf{R-EB} & \textbf{R-TT} & \textbf{DGEA} & \textbf{IKEA} & \textbf{CB} \\
\midrule

\multirow{2}{*}{\makecell[c]{None \\($\text{EE}^{\text{R}}$)}}
& \textbf{Original}
& 20.3 & 56.9 & 11.7 & 60.9 & 24.5 & 26.5 \\
& \textbf{Diversity}
& \textbf{25.1} & \textbf{71.8} & \textbf{12.4} & \textbf{67.2} & \textbf{35.5} & \textbf{27.7} \\
\midrule

\multirow{2}{*}{\makecell[c]{Threshold \\($\text{EE}^{\text{R}}$)}}
& \textbf{Original}
& 11.2 & 28.7 & 6.00 & 24.0 & 31.3 & 22.4 \\
& \textbf{Diversity}
& \textbf{14.3} & \textbf{36.8} & \textbf{7.70} & \textbf{30.4} & \textbf{36.5} & \textbf{23.4} \\
\bottomrule
\end{tabular}
\label{tab-query_diversity}
\end{table}

\subsubsection{Query Diversity Optimization} \label{sec-qd}
Existing attacks encourage diversity primarily by pushing each newly crafted query away from previously extracted chunks; however, they largely overlook redundancy among queries themselves. One can readily envision a trivial case in which all queries remain nearly identical to one another while being maximally distant from the already retrieved knowledge. Such behavior does not yield genuine query diversity and therefore fails to explore distinct regions of the knowledge base.
To address this limitation, we augment all six attack baselines by additionally encouraging each newly generated query to diverge from previously issued queries. This augmentation is uniformly applied to all six attack baselines. We describe implementation details of optimizing query-diversity under different attack paradigms:
\begin{itemize}[leftmargin=*, itemsep=0pt]
    \item \textbf{Explicit Optimization.} 
    For embedding-based attacks such as R-EB and DGEA, we incorporate an additional diversity term into the optimization objective. Concretely, the target embedding for each new query is encouraged via gradient descent to be far from the embeddings of all previously generated queries, thereby explicitly enforcing query-level diversity during optimization.
    \item \textbf{Implicit Optimization.} 
    For attacks without an explicit optimization process, including R-TT, R-TK, IKEA, and CopyBreak, we enforce query diversity through similarity-based filtering. For each attack, a candidate attack query is first generated following its original query generation procedure. We then compute its embedding similarity with all previously issued queries. The candidate is accepted as the next attack query only if its similarity scores fall below a predefined threshold; otherwise, the generation process is repeated until the diversity is satisfied.
\end{itemize}

Table~\ref{tab-query_diversity} reports the average $\text{EE}^{\text{R}}$ under the Original and Query-Diversity settings across four datasets. Query diversity consistently improves extraction effectiveness by enabling broader exploration of the knowledge base. The gains are most pronounced for gradient-based attacks (R-EB and DGEA), where the optimization process directly aligns with the diversity objective. In contrast, R-TT and CB exhibit smaller improvements because their query optimization relies on indirect LLM-based generation and filtering.

\newpage
\subsubsection{Cross-lingual Generalizability} \label{sec-lang}
To evaluate the generalizability of our benchmark beyond English-language datasets, we include two non-English medical datasets: \textbf{Med\_Chinese}~\cite{zeng2020meddialog}, a Chinese patient-doctor dialogue dataset, and \textbf{Med\_Vietnamese}~\cite{vietnamese-medical-qa}, a Vietnamese medical QA dataset. These datasets cover different language families while remaining within the sensitive medical domain. We replace the English retriever with \texttt{multilingual-MiniLM} to accommodate the non-English setting and keep all other settings unchanged. Table~\ref{tab-cross-lang} reports the retrieval extraction effectiveness ($\text{EE}^\text{R}$) and generation extraction effectiveness ($\text{EE}^\text{G}_{\text{LS}}$) on three representative attack and defense settings. Overall, the trends in evaluated settings remain consistent with those in the English domain.
\begin{itemize}[leftmargin=*, itemsep=0pt]
\item \textbf{Summary remains the most effective generation-stage defense}. Across all attacks, it substantially reduces $\text{EE}^\text{G}_{\text{LS}}$. For example, on Med\_Vietnamese, Summary reduces R-TK's $\text{EE}^\text{G}_{\text{LS}}$ from 100.0 to 4.17, demonstrating its effectiveness in preventing verbatim knowledge reproduction across languages.

\item \textbf{Threshold has limited impact on natural-language attacks}. While it significantly reduces DGEA's retrieval effectiveness, its effect on IKEA is minor. This is because IKEA generates semantically natural queries that remain highly relevant to the target knowledge and are therefore unlikely to be filtered by similarity-based defenses.

\item \textbf{IKEA remains stealthy but less effective at generation-stage extraction}. Although IKEA achieves competitive retrieval-stage extraction, its $\text{EE}^\text{G}_{\text{LS}}$ remains close to zero on both datasets. By avoiding explicit extraction-oriented prompts, IKEA is harder to detect but also less capable of inducing the generator to reproduce knowledge base content verbatim. 
\end{itemize}
These results suggest that the attack-defense behaviors identified in our English benchmark are largely language-independent, highlighting the robustness of our benchmark in multilingual settings.

\begin{table}[t!]
\centering
\setlength{\tabcolsep}{2.5pt}
\small
\caption{Cross-lingual evaluation of $\text{EE}^\text{R}$ and $\text{EE}^\text{G}_{\text{LS}}$ on Med\_Chinese and Med\_Vietnamese datasets.}
\vspace{-2ex}
\label{tab:defense_results}
\begin{tabular}{llcccccc}
\toprule
\multirow{2}{*}{\textbf{Dataset}}
&
\multirow{2}{*}{\textbf{Defense}}
&
\multicolumn{2}{c}{\textbf{R-TK}} &
\multicolumn{2}{c}{\textbf{DGEA}} &
\multicolumn{2}{c}{\textbf{IKEA}} \\
\cmidrule(lr){3-4}
\cmidrule(lr){5-6}
\cmidrule(lr){7-8}
&
&
$\text{EE}^{\text{R}}$ &
$\text{EE}^{\text{G}}_{\text{LS}}$ &
$\text{EE}^{\text{R}}$ &
$\text{EE}^{\text{G}}_{\text{LS}}$ &
$\text{EE}^{\text{R}}$ &
$\text{EE}^{\text{G}}_{\text{LS}}$ \\
\midrule

\multirow{3}{*}{\textbf{Med\_Chinese}}
& None
& 15.00 & 60.71
& 51.00 & 53.35
& 15.83 & 0.23 \\

& Threshold 
& 14.83 & 60.92
& 29.00 & 52.76
& 13.67 & 0.24 \\

& Summary
& 15.00 & 36.08
& 51.83 & 20.81
& 15.33 & 0.02 \\

\midrule

\multirow{3}{*}{\textbf{Med\_Vietnamese}}
& None
& 7.50 & 100.00
& 39.50 & 98.99
& 24.17 & 1.98 \\

& Threshold 
& 7.50 & 99.99
& 19.50 & 100.00
& 22.17 & 1.64 \\

& Summary
& 7.50 & 4.17
& 41.00 & 12.88
& 23.50 & 0.11 \\

\bottomrule
\end{tabular}
\label{tab-cross-lang}
\vspace{-5ex}
\end{table}

\subsubsection{Attack Cost Analysis} \label{sec-lang}

\begin{figure}[t!]
    \centering
    \includegraphics[width=\linewidth]{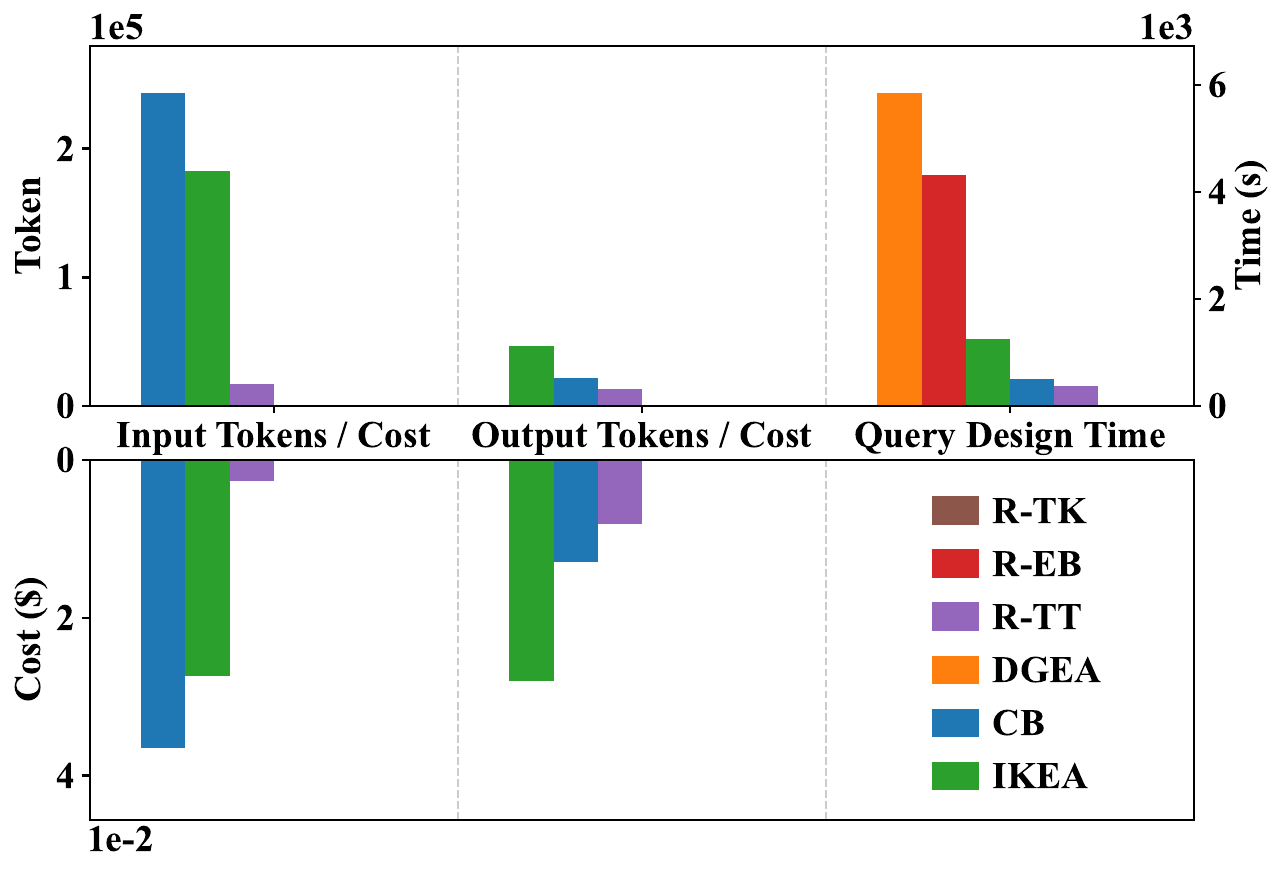}
    \vspace{-4ex}
    \caption{Token and time cost comparison on HarryPotter dataset across different attack methods.}
    \label{fig-tokens}
    \vspace{-4ex}
\end{figure}

We compare the efficiency of different attack baselines in terms of execution time and LLM token consumption. Since LLM providers charge separately for input and output tokens, we report both token consumption and runtime under a fixed query budget of 200. In Figure~\ref{fig-tokens}, DGEA, R-EB, and R-TT do not rely on LLMs for query generation and therefore incur zero token cost. In contrast, IKEA, CopyBreak, and R-TK use LLMs to construct attack queries. Among them, R-TK has the lowest token consumption because it invokes the LLM only once per query to generate a random sentence. IKEA and CopyBreak repeatedly call the LLM during iterative query optimization, resulting in substantially higher token usage. 

Regarding execution time, LLM-based attacks are mainly bottlenecked by LLM inference latency. For non-LLM attacks such as DGEA and R-EB, the runtime is dominated by greedy search over the query space, with a time complexity of $\mathcal{O}(E \times T \times P)$, where $E$ is the number of optimization epochs, $T$ is the query length, and $P$ is the token substitution pool size. DGEA further introduces an embedding optimization step that searches for queries farthest from previously extracted content, making it slower than R-EB. In contrast, R-TK simply samples and concatenates tokens from a predefined pool to form queries, avoiding iterative optimization altogether and thus incurring negligible runtime overhead.

%% file: Conclusion.tex
\section{Conclusion and Future Work}\label{sec-conclusion}
RAG systems are increasingly deployed in high-stakes applications, yet the introduction of external knowledge bases exposes new extraction attack surfaces beyond model parameters and training data. Existing studies adopt heterogeneous experimental settings and model configurations, hindering unified and fair evaluation.
To address this gap, we present the first comprehensive benchmark for knowledge extraction attacks and defenses in RAG systems, unifying the design space and establishing fair, reproducible experimental protocols. Our results show that effective extraction requires optimization at both the retrieval and generation stages. While existing defenses operate at different stages in the RAG pipeline with complementary strengths, no single defense provides complete protection. We further demonstrate that limited query–query diversity leads to redundant exploration, embedding-based attacks exhibit weak cross-model transferability, and both generator instruction-following capabilities and knowledge-base indexing strategies substantially influence extraction vulnerability.
Future work includes multi-level diversity optimization, multi-stage defense coordination, and extending the benchmark to agentic RAG architectures.



%% file: Acknowledgments.tex
\section*{Acknowledgments}
This research is supported by the National Science Foundation (NSF) under grant number IIS 2524379, NAIRR 250188 and NAIRR 260016, and Adobe Research. This research used the Frank cluster at the University or Oregon with the \href{https://e4s.io}{\textcolor{blue}{E4S}} collection of AI tools.

%% file: Appendix.tex

\section{Appendix}





\subsection{Details of Benchmark Baselines}\label{app-baseline}
In this section, we comprehensively review existing extraction baselines~\cite{zeng2024good,cohen2024unleashing,wang2025silent,jiang2025feedback} in our benchmark. Each baseline represents a distinct extraction attack strategy for constructing the \texttt{INFORMATION} component $\mathcal{I}$ of the attack query to steer the retriever toward exploring different embedding regions of the knowledge base. 
These baselines span purely random~\cite{zeng2024good,cohen2024unleashing} and adaptive~\cite{cohen2024unleashing,wang2025silent,jiang2025feedback} methods, thereby covering a broad spectrum of real-world attack behaviors.  
For all baselines, the \texttt{COMMAND} $\mathcal C$ component remains fixed.
\begin{itemize}[leftmargin=*, itemsep=0pt]
    \item \textbf{RandomText (R-TT)}~\cite{jiang2025feedback} attack constructs the $\mathcal I^t$ component by prompting an LLM with a high temperature to produce a syntactically valid yet semantically random natural-language sentence. This allows each attack query to explore diverse regions of the retrieval embedding space without any optimization. 

    \item \textbf{RandomToken (R-TK)} attack constructs \texttt{INFORMATION} at $t^{\text{th}}$ query round $\mathcal I^t$ by concatenating a fixed number of tokens sampled from the vocabulary of the attack embedding model. This provides a simple baseline for embedding-level randomization.

    \item \textbf{RandomEmb (R-EB)}~\cite{cohen2024unleashing} attack begins by collecting a set of English embedding vectors from an external corpus (e.g., WikiText) that is disjoint from the attack-targeted knowledge, thereby preventing information leakage that could make the attack artificially easy. This collection is to estimate an embedding distribution that reflects natural linguistic structure. For the $t^{\text{th}}$ round of attack, a target embedding vector is first sampled from this distribution as a reference. The \texttt{INFORMATION} $\mathcal{I}^t$ is then constructed by initializing a placeholder query and performing greedy token optimization: the algorithm iteratively replaces tokens to maximize the cosine similarity between the evolving query embedding and the sampled target embedding. This procedure enables RandomEmb to explore retrieval embedding space that aligns with natural linguistic structure.

    \item \textbf{Dynamic Greedy Embedding Attack (DGEA)}~\cite{cohen2024unleashing} adaptively constructs \texttt{INFORMATION} $\mathcal{I}^t$ to explore unexplored retrieval embedding regions. At each round, it selects a target embedding farthest from previously extracted chunks and greedily updates $\mathcal I^t$ to maximize the similarity between the resulting query $\mathcal Q^t$ and the target embedding. This iterative process improves retrieval coverage and diversity across the corpus.

    \item \textbf{CopyBreak Attack}~\cite{jiang2025feedback} alternates between \emph{exploration} and \emph{exploitation} to construct $\mathcal I^t$. Exploration generates semantically diverse sentences to probe new embedding regions, while exploitation expands around extracted chunks by generating logically adjacent text from sentence prefixes or suffixes. Repeating these modes at a fixed frequency $N$ enables both broad corpus coverage and local content expansion.

    \item \textbf{Implicit Knowledge Extraction Attack (IKEA)}~\cite{wang2025silent} crafts benign-looking information-seeking queries to evade extraction-intent detection. It first generates a diverse set of domain-relevant anchors and samples one to construct each query. Based on the RAG response, unsuccessful anchors are downweighted while successful anchors and their semantic neighbors are upweighted. By iteratively exploring promising semantic regions until redundancy or blocking occurs, IKEA forms an adaptive, human-like exploration trajectory in the $\mathcal I$-space.
\end{itemize}

\vspace{-2ex}
\subsection{Details of Generator Extraction Metrics}
\label{app:gen-metric-details}
Following Section~\ref{sec-eval-protocol}, we quantify generator extraction effectiveness $\mathrm{EE}^{\mathrm{G}}$ by measuring overlap between the generated output $\mathcal{A}^t$ and the retrieved content $\mathcal{R}^t$. Due to no explicit correspondence between generated responses and retrieved items, we first propose an alignment strategy to pair generated with retrieved content, and then compute similarity for each aligned pair to quantify overlap.

\vspace{-1ex}
\subsubsection{Alignment Strategies}
Depending on attacks, we consider two cases when aligning retrieved with generated contents:
\begin{itemize}[leftmargin=*, itemsep=0pt]
    \item \textbf{Pair-wise Alignment.}
    Attack methods such as DGEA and CopyBreak~\cite{cohen2024unleashing,jiang2025feedback} explicitly include a \texttt{COMMAND} for verbatim leakage. If the retriever returns $k$ knowledge instances at round $t$, the generator outputs $k$ corresponding segments, yielding $k$ well-aligned pairs for evaluation. Metrics can therefore be applied directly on a per-pair basis $\psi_{\text{pair}}(\mathcal A^t, \mathcal R^t)
    = \sum_{i=1}^{|\mathcal R^t|} \psi_{\text{unit}}(\mathcal A^t_i, \mathcal R^t_i)$

    \item \textbf{Concatenated Alignment.}
    In contrast, IKEA~\cite{wang2025silent} does not use a \texttt{COMMAND} component, and therefore the RAG generator produces a single paragraph-style response that blends information across all retrieved knowledge instances, preventing a one-to-one alignment. In this case, we concatenate all retrieved knowledge instances into a single reference text and compute alignment as:
    $\tilde{\mathcal R}^t = \text{concat}(\mathcal R^t_1, \ldots, \mathcal R^t_{|\mathcal R^t|}), \psi_{\text{pair}}(\mathcal A^t, \mathcal R^t)
    = \psi_{\text{unit}}(\mathcal A^t, \tilde{\mathcal R}^t)$
    If the generator refuses to answer (e.g., outputs a refusal or safety message), the corresponding alignment score is set to $0$.
\end{itemize}

\subsubsection{Similarity Instantiations}

The unit-level alignment function $\psi_{\text{unit}}$ can be instantiated from semantic and lexical perspectives:

\begin{itemize}[leftmargin=*, itemsep=0pt]
    \item \textbf{Lexical Similarity}
    evaluates extraction at the lexical level. Common instantiations include Exact Match~\cite{rajpurkar2016squad}, BLEU~\cite{papineni2002bleu}, ROUGE-L~\cite{lin2004rouge}, which compare the token-level overlap between generated output and the retrieved one. High lexical similarity indicates that the generator reproduced the retrieved content in a nearly verbatim manner. In this work, we use ROUGE-L following~\cite{zeng2024good,wang2025silent,jiang2025feedback}.

    \item \textbf{Semantic Similarity.}
    Semantic similarity evaluates extraction at the semantic level using embedding-based similarity measures. A common instantiation~\cite{wang2025silent,jiang2025feedback} is cosine similarity between embeddings of the generated output and the retrieved one. High semantic similarity indicates that the generator conveys information that is close in meaning to the retrieved content.
\end{itemize}

\subsection{Knowledge Base Setup} \label{app-indexing}

We investigate how different RAG indexing strategies affect attack performance. Specifically, we compare three representative indexings: \textbf{Instance indexing}, where each index entry corresponds to a natural data instance; \textbf{Fixed-Chunk indexing}, which segments the knowledge base into fixed-length text chunks with 20\% overlap~\cite{jiang2025feedback}; and \textbf{Graph Triplet indexing}~\cite{Liu_LlamaIndex_2022}, which transforms document sentences into structured triplets of entity-relation-entity.

Evaluating attacks across different indexing strategies presents non-trivial challenges. Raw item-level leakage counts are not comparable across indexings because the granularity and semantic content of stored items differ substantially. For example, Graph indexing produces many fine-grained triplets while chunk-based indexing produces fewer but more information-dense text chunks.

To address this issue, we adopt a target-oriented evaluation strategy for the numerator of our metric. Instead of counting how many indexed items are leaked, we measure how much \emph{key private information} is extracted, formalized as $\phi\left(\cup_{t=1}^{T} \mathcal{R}^t, \mathcal{D}^*\right)$, where $\mathcal{R}^t$ represents the retrieved content at query $t$ and $\mathcal{D}^*$ denotes the set of key private information units shared across all indexing strategies. This design enables fair comparison by anchoring evaluation to semantic targets rather than indexing artifacts.

In addition, different indexing strategies retrieve items with varying amounts of information. Text chunks may include a large number of non-informative tokens (e.g., stop words), whereas GraphRAG retrieval tends to return concise, content-dense triplets. To mitigate this discrepancy, we introduce token-length normalization for the denominator, computed as $\sum_{t=1}^{T}|\mathcal{R}^t|_{\text{token}}$, which sums the total number of tokens retrieved across all $T$ attack queries. This normalization strategy prevents biases caused by differences in textual verbosity across indexing methods.

To conclude, our new evaluation metric is:
\begin{equation}
    \text{EE}^{\text{R}}_{\text{token}} = 
    \phi\left(\cup_{t=1}^{T} \mathcal{R}^t, \mathcal{D}^*\right)(\sum_{t=1}^{T}|\mathcal{R}^t|_{\text{token}})^{-1}
\end{equation}

\begin{table}[t]
\centering
\small
\setlength{\tabcolsep}{4.5pt}
\caption{SAGE defense cost analysis per dataset, total token consumption, estimated cost, and processing time.}
\vspace{-2ex}
\label{tab:dataset_stats}
\begin{tabular}{lcccc}
\toprule
\textbf{Dataset} &
\textbf{Instances} &
\textbf{Total Tokens} &
\textbf{Total Cost} &
\textbf{Time} \\
\midrule

\textbf{HealthCareMagic} & 112K & 856.4M & \$217.5 & $>$10d \\
\textbf{Enron}           & 517K & 22.6B  & \$6,096.8 & $>$10d \\
\textbf{HarryPotter}     & 26K  & 137.1M & \$33.6 & $\sim$30h \\
\textbf{Pokémon}         & 1.3K & 6.6M   & \$1.6 & $\sim$2h \\
\bottomrule
\end{tabular}
\label{tab-sage-cost}
\end{table}

\begin{figure}[t!]
    \centering
    \includegraphics[width=\linewidth]{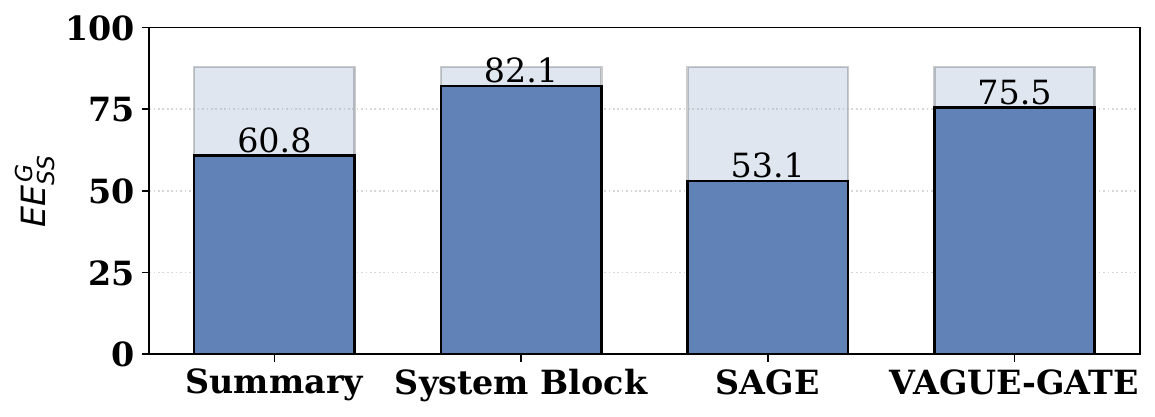}
    \vspace{-5ex}
    \caption{Average $\text{EE}^{\text{G}}_{\text{SS}}$ across HarryPotter and Pokémon for all defenses across all attacks.}
    \label{fig-rewrite}
\end{figure}

\subsection{Rewrite-Based Defense Analysis}

Beyond the defenses evaluated in the main paper, we additionally study two representative rewrite-based defenses, SAGE~\cite{zeng2025mitigating} and VAGUE-GATE~\cite{hemmat2025vague}. SAGE operates at indexing time by replacing the original knowledge base with a rewritten corpus that preserves the semantic content of the source documents while avoiding verbatim text reproduction. Retrieval is then performed exclusively over this rewritten corpus, preventing direct access to the original documents. In contrast, VAGUE-GATE operates at generation time by identifying sensitive content within retrieved passages and selectively paraphrasing it before the passages are provided to the generator. This design aims to reduce the risk of memorization and verbatim leakage while preserving the utility of the retrieved information. 
We evaluate these defenses in conjunction with Summary and System Block, reporting the average generation-stage extraction effectiveness ($\text{EE}^{\text{G}}_{\text{SS}}$) over all attacks on the Harry Potter and Pokémon datasets. Our evaluation is restricted to these datasets because SAGE necessitates offline rewriting of the complete corpus, resulting in substantial computational overhead that renders experiments on larger-scale datasets prohibitively expensive (Table~\ref{tab-sage-cost}).

Figure~\ref{fig-rewrite} shows that \textbf{SAGE is the most effective generation-stage defense}, reducing average $\text{EE}^{\text{G}}_{\text{SS}}$ to 53.1 and outperforming both Summary and System Block. In contrast, \textbf{VAGUE-GATE provides only moderate protection}. Since it primarily perturbs sensitive entities while preserving the surrounding context, it leaves much of the surrounding context unchanged and preserves high semantic similarity to the original content.

However, rewrite-based defenses incur substantial deployment costs. As in Table~\ref{tab-sage-cost}, applying SAGE to large corpora such as HealthCareMagic and Enron would require rewriting approximately 856M and 22B tokens, respectively, corresponding to estimated costs of \$217 and \$6,097 and more than ten days of processing time.

Overall, these results reveal a clear trade-off: rewrite-based defenses, especially SAGE, significantly reduce knowledge extraction attack effectiveness by modifying the corpus itself, but their high preprocessing cost limits practicality for large-scale deployments.







%% file: others.bib
@inproceedings{survey2025pii,
  author = {Cheng, Shuai and Li, Zhao and Meng, Shu and Ren, Mengxia and Xu, Haitao and Hao, Shuai and Yue, Chuan and Zhang, Fan},
  title = {{Understanding PII Leakage in Large Language Models: A Systematic Survey}},
  booktitle = {Proceedings of the 34th International Joint Conference on Artificial Intelligence (IJCAI-25), Survey Track},
  year = {2025},
  pages = {10409--10416}
}

@misc{li2025confidential,
      title={Confidential Prompting: Privacy-preserving LLM Inference on Cloud}, 
      author={Caihua Li and In Gim and Lin Zhong},
      year={2025},
      eprint={2409.19134},
      archivePrefix={arXiv},
      primaryClass={cs.CR},
      url={https://arxiv.org/abs/2409.19134}, 
}

@inproceedings{gonzalez2021user,
  title={User profiling by network observers},
  author={Gonzalez, Roberto and Soriente, Claudio and Carrascosa, Juan Miguel and Garcia-Duran, Alberto and Iordanou, Costas and Niepert, Mathias},
  booktitle={Proceedings of the 17th International Conference on emerging Networking EXperiments and Technologies},
  pages={212--222},
  year={2021}
}

@inproceedings{he2023understanding,
  title={Understanding and mitigating hardware failures in deep learning training systems},
  author={He, Yi and Hutton, Mike and Chan, Steven and De Gruijl, Robert and Govindaraju, Rama and Patil, Nishant and Li, Yanjing},
  booktitle={Proceedings of the 50th Annual International Symposium on Computer Architecture},
  pages={1--16},
  year={2023}
}

@inproceedings{liang2025attnchecker,
  title={ATTNChecker: Highly-Optimized Fault Tolerant Attention for Large Language Model Training},
  author={Liang, Yuhang and Li, Xinyi and Ren, Jie and Li, Ang and Fang, Bo and Chen, Jieyang},
  booktitle={Proceedings of the 30th ACM SIGPLAN Annual Symposium on Principles and Practice of Parallel Programming},
  pages={252--266},
  year={2025}
}

@article{xu2025comprehensive,
  title={A Comprehensive Survey of Agentic AI in Healthcare},
  author={Xu, Gelei and Li, Xueyang and Chen, Yixiong and Duan, Yuying and Wu, Shuqing and Yu, Alexander and Chiu, Ching-Hao and Ni, Juntong and Tang, Ningzhi and Li, Toby Jia-Jun and others},
  journal={Authorea Preprints},
  year={2025},
  publisher={Authorea}
}


%% file: reference.bib
@article{zeng2024good,
  title={The good and the bad: Exploring privacy issues in retrieval-augmented generation (rag)},
  author={Zeng, Shenglai and Zhang, Jiankun and He, Pengfei and Xing, Yue and Liu, Yiding and Xu, Han and Ren, Jie and Wang, Shuaiqiang and Yin, Dawei and Chang, Yi and others},
  journal={arXiv preprint arXiv:2402.16893},
  year={2024}
}

@article{cohen2024unleashing,
  title={Unleashing worms and extracting data: Escalating the outcome of attacks against rag-based inference in scale and severity using jailbreaking},
  author={Cohen, Stav and Bitton, Ron and Nassi, Ben},
  journal={arXiv preprint arXiv:2409.08045},
  year={2024}
}

@article{wang2025silent,
  title={Silent leaks: Implicit knowledge extraction attack on rag systems through benign queries},
  author={Wang, Yuhao and Qu, Wenjie and Zhai, Shengfang and Jiang, Yanze and Liu, Zichen and Liu, Yue and Dong, Yinpeng and Zhang, Jiaheng},
  journal={arXiv preprint},
  year={2025}
}

@article{jiang2025feedback,
  title={Feedback-guided extraction of knowledge base from retrieval-augmented LLM applications},
  author={Jiang, Changyue and Pan, Xudong and Hong, Geng and Bao, Chenfu and Yang, Min},
  journal={arXiv preprint arXiv:2411.14110},
  year={2025}
}

@article{lewis2020retrieval,
  title={Retrieval-augmented generation for knowledge-intensive nlp tasks},
  author={Lewis, Patrick and Perez, Ethan and Piktus, Aleksandra and Petroni, Fabio and Karpukhin, Vladimir and Goyal, Naman and K{\"u}ttler, Heinrich and Lewis, Mike and Yih, Wen-tau and Rockt{\"a}schel, Tim and others},
  journal={Advances in Neural Information Processing Systems},
  volume={33},
  pages={9459--9474},
  year={2020}
}

@misc{liu2022llama,
  author = {Liu, Jerry},
  title = {LlamaIndex},
  howpublished = {11 2022. \url{https://github.com/jerryjliu/llama_index}},
  year = {2022}
}

@misc{chase2022langchain,
  author = {Chase, Harrison},
  title = {LangChain},
  howpublished = {October 2022. \url{https://github.com/hwchase17/langchain}},
  year = {2022}
}

@article{van2023clinical,
  title={Clinical text summarization: Adapting large language models can outperform human experts},
  author={Van Veen, Dave and Van Uden, Cara and Blankemeier, Louis and Delbrouck, Jean-Benoit and Aali, Asad and Bluethgen, Christian and Pareek, Anuj and Polacin, Malgorzata and Collins, William and Ahuja, Neera and others},
  journal={arXiv preprint arXiv:2309.07430},
  year={2023}
}

@article{ram2023context,
  title={In-context retrieval-augmented language models},
  author={Ram, Ori and Levine, Yoav and Dalmedigos, Itay and Muhlgay, Dor and Shashua, Amnon and Leyton-Brown, Kevin and Shoham, Yoav},
  journal={arXiv preprint arXiv:2302.00083},
  year={2023}
}

@misc{lavita_chatdoctor_healthcaremagic_2023,
  title        = {ChatDoctor-HealthCareMagic-100k},
  author       = {lavita},
  howpublished = {\url{https://huggingface.co/datasets/lavita/ChatDoctor-HealthCareMagic-100k}},
  year         = {2023}
}

@inproceedings{klimt2004enron,
  title        = {The Enron Corpus: A New Dataset for Email Classification Research},
  author       = {Klimt, Bryan and Yang, Yiming},
  booktitle    = {European Conference on Machine Learning},
  pages        = {217--226},
  year         = {2004},
  organization = {Springer}
}

@misc{vapit_harrypotterqa_2023,
  title        = {HarryPotterQA},
  author       = {vapit},
  howpublished = {\url{https://huggingface.co/datasets/vapit/HarryPotterQA}},
  year         = {2023}
}

@misc{tungdop2_pokemon_2023,
  title        = {Pok\'emon QA Dataset},
  author       = {Duong, Quang Tung},
  howpublished = {\url{https://huggingface.co/datasets/tungdop2/pokemon}},
  year         = {2023}
}

@article{tan2025equilibrate,
  title={Equilibrate rlhf: Towards balancing helpfulness-safety trade-off in large language models},
  author={Tan, Yingshui and Jiang, Yilei and Li, Yanshi and Liu, Jiaheng and Bu, Xingyuan and Su, Wenbo and Yue, Xiangyu and Zhu, Xiaoyong and Zheng, Bo},
  journal={arXiv preprint arXiv:2502.11555},
  year={2025}
}

@article{zhang2021emailsum,
  title={EmailSum: Abstractive email thread summarization},
  author={Zhang, Shiyue and Celikyilmaz, Asli and Gao, Jianfeng and Bansal, Mohit},
  journal={arXiv preprint arXiv:2107.14691},
  year={2021}
}

@article{hearst1997text,
  title={Text tiling: Segmenting text into multi-paragraph subtopic passages},
  author={Hearst, Marti A},
  journal={Computational linguistics},
  volume={23},
  number={1},
  pages={33--64},
  year={1997}
}

@article{liu2025exposing,
  title={Exposing Privacy Risks in Graph Retrieval-Augmented Generation},
  author={Liu, Jiale and Zhang, Jiahao and Wang, Suhang},
  journal={arXiv preprint arXiv:2508.17222},
  year={2025}
}

@article{wei2023jailbroken,
  title={Jailbroken: How does llm safety training fail?},
  author={Wei, Alexander and Haghtalab, Nika and Steinhardt, Jacob},
  journal={Advances in Neural Information Processing Systems},
  volume={36},
  pages={80079--80110},
  year={2023}
}

@inproceedings{papineni2002bleu,
  title={Bleu: a method for automatic evaluation of machine translation},
  author={Papineni, Kishore and Roukos, Salim and Ward, Todd and Zhu, Wei-Jing},
  booktitle={Proceedings of the 40th annual meeting of the Association for Computational Linguistics},
  pages={311--318},
  year={2002}
}

@inproceedings{lin2004rouge,
  title={Rouge: A package for automatic evaluation of summaries},
  author={Lin, Chin-Yew},
  booktitle={Text summarization branches out},
  pages={74--81},
  year={2004}
}

@article{rajpurkar2016squad,
  title={Squad: 100,000+ questions for machine comprehension of text},
  author={Rajpurkar, Pranav and Zhang, Jian and Lopyrev, Konstantin and Liang, Percy},
  journal={arXiv preprint},
  year={2016}
}

@article{qi2024follow,
  title={Follow my instruction and spill the beans: Scalable data extraction from retrieval-augmented generation systems},
  author={Qi, Zhenting and Zhang, Hanlin and Xing, Eric and Kakade, Sham and Lakkaraju, Himabindu},
  journal={arXiv preprint arXiv:2402.17840},
  year={2024}
}

@article{wang2020minilm,
  title={Minilm: Deep self-attention distillation for task-agnostic compression of pre-trained transformers},
  author={Wang, Wenhui and Wei, Furu and Dong, Li and Bao, Hangbo and Yang, Nan and Zhou, Ming},
  journal={Advances in neural information processing systems},
  year={2020}
}

@article{li2023towards,
  title={Towards general text embeddings with multi-stage contrastive learning},
  author={Li, Zehan and Zhang, Xin and Zhang, Yanzhao and Long, Dingkun and Xie, Pengjun and Zhang, Meishan},
  journal={arXiv preprint},
  year={2023}
}

@article{chen2024bge,
  title={Bge m3-embedding: Multi-lingual, multi-functionality, multi-granularity text embeddings through self-knowledge distillation},
  author={Chen, Jianlv and Xiao, Shitao and Zhang, Peitian and Luo, Kun and Lian, Defu and Liu, Zheng},
  journal={arXiv preprint},
  year={2024}
}

@inproceedings{zhang2025intention,
  title={Intention analysis makes llms a good jailbreak defender},
  author={Zhang, Yuqi and Ding, Liang and Zhang, Lefei and Tao, Dacheng},
  booktitle={Proceedings of the 31st International Conference on Computational Linguistics},
  year={2025}
}

@article{merity2016pointer,
  title={Pointer sentinel mixture models},
  author={Merity, Stephen and Xiong, Caiming and Bradbury, James and Socher, Richard},
  journal={arXiv preprint arXiv:1609.07843},
  year={2016}
}

@article{shi2023replug,
  title={Replug: Retrieval-augmented black-box language models},
  author={Shi, Weijia and Min, Sewon and Yasunaga, Michihiro and Seo, Minjoon and James, Rich and Lewis, Mike and Zettlemoyer, Luke and Yih, Wen-tau},
  journal={arXiv preprint arXiv:2301.12652},
  year={2023}
}

@article{russo2024face,
  title={Face the Facts! Evaluating RAG-based Fact-checking Pipelines in Realistic Settings},
  author={Russo, Daniel and Menini, Stefano and Staiano, Jacopo and Guerini, Marco},
  journal={arXiv preprint arXiv:2412.15189},
  year={2024}
}

@article{li2022survey,
  title={A survey on retrieval-augmented text generation},
  author={Li, Huayang and Su, Yixuan and Cai, Deng and Wang, Yan and Liu, Lemao},
  journal={arXiv preprint arXiv:2202.01110},
  year={2022}
}

@inproceedings{carlini2021extracting,
  title={Extracting training data from large language models},
  author={Carlini, Nicholas and Tramer, Florian and Wallace, Eric and Jagielski, Matthew and Herbert-Voss, Ariel and Lee, Katherine and Roberts, Adam and Brown, Tom and Song, Dawn and Erlingsson, Ulfar and others},
  booktitle={30th USENIX Security Symposium (USENIX Security 21)},
  pages={2633--2650},
  year={2021}
}

@inproceedings{kandpal2022deduplicating,
  title={Deduplicating training data mitigates privacy risks in language models},
  author={Kandpal, Nikhil and Wallace, Eric and Raffel, Colin},
  booktitle={International Conference on Machine Learning},
  pages={10697--10707},
  year={2022},
  organization={PMLR}
}

@article{carlini2022quantifying,
  title={Quantifying memorization across neural language models},
  author={Carlini, Nicholas and Ippolito, Daphne and Jagielski, Matthew and Lee, Katherine and Tramer, Florian and Zhang, Chiyuan},
  journal={arXiv preprint arXiv:2202.07646},
  year={2022}
}

@article{zeng2023exploring,
  title={Exploring Memorization in Fine-tuned Language Models},
  author={Zeng, Shenglai and Li, Yaxin and Ren, Jie and Liu, Yiding and Xu, Han and He, Pengfei and Xing, Yue and Wang, Shuaiqiang and Tang, Jiliang and Yin, Dawei},
  journal={arXiv preprint arXiv:2310.06714},
  year={2023}
}

@inproceedings{liang2024model,
  title={Model extraction attacks revisited},
  author={Liang, Jiacheng and Pang, Ren and Li, Changjiang and Wang, Ting},
  booktitle={Proceedings of the 19th ACM Asia Conference on Computer and Communications Security},
  pages={1231--1245},
  year={2024}
}

@article{alam2025astuterag,
  title={AstuteRAG-FQA: Task-Aware Retrieval-Augmented Generation Framework for Proprietary Data Challenges in Financial Question Answering},
  author={Alam, Mohammad Zahangir and Zaman, Khandoker Ashik Uz and Miraz, Mahdi H},
  journal={arXiv preprint arXiv:2510.27537},
  year={2025}
}

@article{shi2025hypercube,
  title={Hypercube-RAG: Hypercube-Based Retrieval-Augmented Generation for In-domain Scientific Question-Answering},
  author={Shi, Jimeng and Zhou, Sizhe and Jin, Bowen and Hu, Wei and Wang, Shaowen and Narasimhan, Giri and Han, Jiawei},
  journal={arXiv preprint},
  year={2025}
}

@article{zhang2025benchmarking,
  title={Benchmarking Poisoning Attacks against Retrieval-Augmented Generation},
  author={Zhang, Baolei and Xin, Haoran and Li, Jiatong and Zhang, Dongzhe and Fang, Minghong and Liu, Zhuqing and Nie, Lihai and Liu, Zheli},
  journal={arXiv preprint arXiv:2505.18543},
  year={2025}
}

@article{zhang2024adversarial,
  title={Adversarial hubness in multi-modal retrieval},
  author={Zhang, Tingwei and Suya, Fnu and Jha, Rishi and Zhang, Collin and Shmatikov, Vitaly},
  journal={arXiv preprint arXiv:2412.14113},
  year={2024}
}

@inproceedings{zou2025poisonedrag,
  title={$\{$PoisonedRAG$\}$: Knowledge corruption attacks to $\{$Retrieval-Augmented$\}$ generation of large language models},
  author={Zou, Wei and Geng, Runpeng and Wang, Binghui and Jia, Jinyuan},
  booktitle={34th USENIX Security Symposium (USENIX Security 25)},
  pages={3827--3844},
  year={2025}
}

@article{gao2023retrieval,
  title={Retrieval-augmented generation for large language models: A survey},
  author={Gao, Yunfan and Xiong, Yun and Gao, Xinyu and Jia, Kangxiang and Pan, Jinliu and Bi, Yuxi and Dai, Yixin and Sun, Jiawei and Wang, Haofen and Wang, Haofen},
  journal={arXiv preprint arXiv:2312.10997},
  volume={2},
  number={1},
  year={2023}
}

@article{wang2024knowledge,
  title={Knowledge editing for large language models: A survey},
  author={Wang, Song and Zhu, Yaochen and Liu, Haochen and Zheng, Zaiyi and Chen, Chen and Li, Jundong},
  journal={ACM Computing Surveys},
  volume={57},
  number={3},
  pages={1--37},
  year={2024},
  publisher={ACM New York, NY}
}

@article{rahman2025generative,
  title={Generative AI for Advanced Cyber Defense},
  author={Rahman, Moqsadur and Sanchez, Aaron and Piryani, Krish and Das, Siddhartha and Munikoti, Sai and de la Torre Quintana, Luis and Hasan, Monowar and Aguayo, Joseph and Akbar, Monika and Hossain, Shahriar and others},
  journal={AI for Cybersecurity: Research and Practice},
  pages={109--146},
  year={2025},
  publisher={Wiley Online Library}
}

@techreport{rahman2024retrieval,
  title={Retrieval augmented generation for robust cyber defense},
  author={Rahman, Moqsadur and Piryani, Krish O and Sanchez, Aaron M and Munikoti, Sai and De La Torre, Luis and Levin, Maxwell S and Akbar, Monika and Hossain, Mahmud and Hasan, Monowar and Halappanavar, Mahantesh},
  year={2024},
  institution={Pacific Northwest National Laboratory (PNNL), Richland, WA (United States)}
}

@article{zhangpersonalization,
  title={Personalization of Large Language Models: A Survey},
  author={Zhang, Zhehao and Rossi, Ryan A and Kveton, Branislav and Shao, Yijia and Yang, Diyi and Zamani, Hamed and Dernoncourt, Franck and Barrow, Joe and Yu, Tong and Kim, Sungchul and others}
}

@article{sahu2025knowledge,
  title={Knowledge Homophily in Large Language Models},
  author={Sahu, Utkarsh and Qi, Zhisheng and Halappanavar, Mahantesh and Lipka, Nedim and Rossi, Ryan A and Dernoncourt, Franck and Zhang, Yu and Ma, Yao and Wang, Yu},
  journal={arXiv preprint arXiv:2509.23773},
  year={2025}
}

@article{wu2025retrieval,
  title={Retrieval augmented generation-driven information retrieval and question answering in construction management},
  author={Wu, Chengke and Ding, Wenjun and Jin, Qisen and Jiang, Junjie and Jiang, Rui and Xiao, Qinge and Liao, Longhui and Li, Xiao},
  journal={Advanced Engineering Informatics},
  volume={65},
  pages={103158},
  year={2025},
  publisher={Elsevier}
}

@article{han2024retrieval,
  title={Retrieval-augmented generation with graphs (graphrag)},
  author={Han, Haoyu and Wang, Yu and Shomer, Harry and Guo, Kai and Ding, Jiayuan and Lei, Yongjia and Halappanavar, Mahantesh and Rossi, Ryan A and Mukherjee, Subhabrata and Tang, Xianfeng and others},
  journal={arXiv preprint arXiv:2501.00309},
  year={2024}
}

@article{ling2025domain,
  title={Domain specialization as the key to make large language models disruptive: A comprehensive survey},
  author={Ling, Chen and Zhao, Xujiang and Lu, Jiaying and Deng, Chengyuan and Zheng, Can and Wang, Junxiang and Chowdhury, Tanmoy and Li, Yun and Cui, Hejie and Zhang, Xuchao and others},
  journal={ACM Computing Surveys},
  volume={58},
  number={3},
  pages={1--39},
  year={2025},
  publisher={ACM New York, NY}
}

@article{singh2025agentic,
  title={Agentic retrieval-augmented generation: A survey on agentic rag},
  author={Singh, Aditi and Ehtesham, Abul and Kumar, Saket and Khoei, Tala Talaei},
  journal={arXiv preprint arXiv:2501.09136},
  year={2025}
}

@article{xu2025mem,
  title={A-mem: Agentic memory for llm agents},
  author={Xu, Wujiang and Liang, Zujie and Mei, Kai and Gao, Hang and Tan, Juntao and Zhang, Yongfeng},
  journal={arXiv preprint arXiv:2502.12110},
  year={2025}
}

@article{zeng2024structural,
  title={On the structural memory of llm agents},
  author={Zeng, Ruihong and Fang, Jinyuan and Liu, Siwei and Meng, Zaiqiao},
  journal={arXiv preprint arXiv:2412.15266},
  year={2024}
}

@article{sapkota2025ai,
  title={Ai agents vs. agentic ai: A conceptual taxonomy, applications and challenges},
  author={Sapkota, Ranjan and Roumeliotis, Konstantinos I and Karkee, Manoj},
  journal={arXiv preprint arXiv:2505.10468},
  year={2025}
}

@article{mukhopadhyay2025privacybench,
  title={PrivacyBench: A Conversational Benchmark for Evaluating Privacy in Personalized AI},
  author={Mukhopadhyay, Srija and Reddy, Sathwik and Muthukumar, Shruthi and An, Jisun and Kumaraguru, Ponnurangam},
  journal={arXiv preprint arXiv:2512.24848},
  year={2025}
}

@inproceedings{chandrasekaran2020exploring,
  title={Exploring connections between active learning and model extraction},
  author={Chandrasekaran, Varun and Chaudhuri, Kamalika and Giacomelli, Irene and Jha, Somesh and Yan, Songbai},
  booktitle={29th USENIX Security Symposium (USENIX Security 20)},
  year={2020}
}

@software{Liu_LlamaIndex_2022,
author = {Liu, Jerry},
doi = {10.5281/zenodo.1234},
month = {11},
title = {{LlamaIndex}},
url = {https://github.com/jerryjliu/llama_index},
year = {2022}
}

@inproceedings{zeng2020meddialog,
  title={MedDialog: Large-scale medical dialogue datasets},
  author={Zeng, Guangtao and Yang, Wenmian and Ju, Zeqian and Yang, Yue and Wang, Sicheng and Zhang, Ruisi and Zhou, Meng and Zeng, Jiaqi and Dong, Xiangyu and Zhang, Ruoyu and others},
  booktitle={Proceedings of the 2020 conference on empirical methods in natural language processing (EMNLP)},
  year={2020}
}

@misc{vietnamese-medical-qa,
      title={Vietnamese Medical QA: Question Answering dataset for medical in Vietnamese},
      author={Hung Nguyen},
      year={2024},
}

@inproceedings{zeng2025mitigating,
  title={Mitigating the privacy issues in retrieval-augmented generation (rag) via pure synthetic data},
  author={Zeng, Shenglai and Zhang, Jiankun and He, Pengfei and Ren, Jie and Zheng, Tianqi and Lu, Hanqing and Xu, Han and Liu, Hui and Xing, Yue and Tang, Jiliang},
  booktitle={Proceedings of the 2025 Conference on Empirical Methods in Natural Language Processing},
  year={2025}
}

@inproceedings{hemmat2025vague,
  title={VAGUE-Gate: Plug-and-Play Local-Privacy Shield for Retrieval-Augmented Generation},
  author={Hemmat, Arshia and Moqadas, Matin and Mamanpoosh, Ali and Rismanchian, Amirmasoud and Fatemi, Afsaneh},
  booktitle={Proceedings of the 14th International Joint Conference on Natural Language Processing and the 4th Conference of the Asia-Pacific Chapter of the Association for Computational Linguistics},
  pages={3715--3730},
  year={2025}
}

@article{han2025rag,
  title={Rag vs. graphrag: A systematic evaluation and key insights},
  author={Han, Haoyu and Ma, Li and Wang, Yu and Shomer, Harry and Lei, Yongjia and Qi, Zhisheng and Guo, Kai and Hua, Zhigang and Long, Bo and Liu, Hui and others},
  journal={arXiv preprint arXiv:2502.11371},
  year={2025}
}
